\documentclass[11pt]{article}

\parskip=\medskipamount

\usepackage{amsthm,amsmath,latexsym,amssymb,amsfonts,amscd,mathrsfs}
\usepackage{graphics,lscape,fancyhdr,array,stmaryrd,euscript}
\usepackage{empheq,slashed}
\usepackage{sidecap}
\usepackage{ifpdf}
\usepackage{verbatim}
\usepackage{color}
\usepackage{relsize}
\usepackage{tikz-cd}
\numberwithin{equation}{section}
\usepackage[pdftex,
            colorlinks=true,
            pdfstartview=FitV,
            urlcolor=ColorURL,
            linkcolor=ColorLink]
            {hyperref}
\usepackage{setspace}
\usepackage[numbers,sort&compress]{natbib}
\usepackage[nottoc]{tocbibind}

\usepackage[dvipsnames]{xcolor}
\definecolor{arXiv}{named}{Maroon}
\definecolor{ColorURL}{named}{RoyalBlue}
\definecolor{ColorLink}{named}{Black}
\newcommand{\pl}{\partial}

\newcommand{\inv}{{\scriptscriptstyle-1}}

\newcommand{\lrangle}[1]{\langle #1 \rangle}

\begin{document}
%%%%%%%%%%%%%%%%%%%%%%%%%%%%%%%%%%%%%%%%%%%%%%%%%%%%%%%%%%

\pagenumbering{gobble}
\hfill
\vskip 0.05\textheight
\begin{center}

{\Large\bfseries 
Spinor-helicity formalism for continuous-spin particles}

\vspace{0.4cm}

\vskip 0.03\textheight
\renewcommand{\thefootnote}{\fnsymbol{footnote}}
Thomas \textsc{Basile}${}^{a}$, 
Xavier \textsc{Bekaert}${}^{b}$, 
Felipe \textsc{Figueroa}${}^{a}$ 
\& Evgeny \textsc {Skvortsov}\footnote{Also at Lebedev Institute of Physics, Moscow, Russia}${}^{a}$
\renewcommand{\thefootnote}{\arabic{footnote}}
\vskip 0.03\textheight

{\em ${}^{a}$ Service de Physique de l'Univers, Champs et Gravitation, \\ Universit\'e de Mons, 20 place du Parc, 7000 Mons, 
Belgium}\\
\vspace*{5pt}
{\em ${}^{b}$ Institut Denis Poisson, Unit\'e Mixte de Recherche $7013$ du CNRS,\protect\\ 
Universit\'e de Tours \& Universit\'e d'Orl\'eans,
Parc de Grandmont, 37200 Tours, France}\\
\end{center}

\vskip 0.02\textheight

\begin{abstract}
    We propose a new formulation of continuous-spin particles (CSP) with the help of a single two-component spinor to build asymptotic states. This formulation allows to write down amplitudes in a straightforward way, similar to the massless spinor-helicity approach. The helicity states naturally emerge upon decomposing into components of a fixed homogeneity degree. We show that CSP amplitudes can be understood as the infinite-spin limit of amplitudes of massive particles and, similarly to the scattering of black holes, a certain universal factor exponentiates. In analogy with the recent discovery of collinear amplitudes in self-dual theories, we find nontrivial CSP collinear amplitudes where the collinear fractions are constrained by the dimensionful parameter characterising the various continuous-spin particles.
\end{abstract}

\newpage
\pagenumbering{arabic}

\setcounter{footnote}{0}
\tableofcontents
\newpage

%%%%%%%%%%%%%%%%%%%%%%
\section{Introduction}
%%%%%%%%%%%%%%%%%%%%%%

Ever since the classification by Wigner of all unitary irreducible representations (UIRs) of the Poincar\'e group \cite{Wigner:1939cj}, the appearance of massless representations with infinitely-many degrees of freedom per spacetime point has intrigued theoretical physicists (see e.g. \cite{Bekaert:2017khg} for a review on this topic). They are usually referred to as
``continuous-spin'' representations because they are labelled by a (dimensionful) continuous parameter. Yet, this name is somewhat misleading since they have a discrete spectrum of helicities, more precisely, all integer helicities for the bosonic representation or all half-integer helicities for the fermionic one. Accordingly, they are sometimes called  ``infinite spin'' representations to stress the latter fact.

Various people (notably Wigner himself) have proposed to discard them as ``unphysical'' based on various no-go arguments concluding, e.g. that their infinite spin would imply an infinite heat capacity \cite{Wigner:1963wwt} (see, however, \cite{Schuster:2024wjc}), that they cannot be realized as local quantum fields \cite{Yngvason:1970fy,Iverson:1971hq,Abbott:1976bb,Hirata:1977ss,Schroer:1996rx,Mund:2004sy,Longo:2015tra,Kohler:2015cua,Schroer:2015rct} (nonetheless, they admit local field equations, see e.g. \cite{Wigner1947,Bargmann:1948ck,Bekaert:2005in}), etc. 
Nevertheless, the last decade experienced a growing surge of interest in these unusual massless representations.
In fact, the weakness of any no-go theorem is the strength of its underlying hypotheses. The crucial feature of continuous-spin fields is that they have an infinite number of degrees of freedom (per spacetime point) hence they sit at the frontier of traditional Quantum Field Theory (QFT) and their proper treatment calls for new formalisms, suited for this exotic representation. There is a nontrivial overlap with the idea of higher-spin gravities, see e.g. \cite{Bekaert:2022poo}, since in the limit of vanishing ``internal mass'' continuous-spin particles reduce to a direct sum of massless particles of all spins.

Several formulations of continuous-spin fields have already been proposed, leading to interesting results. For instance, the original Wigner equations \cite{Wigner1947,Bargmann:1948ck} admit an equivalent gauge formulation \cite{Bekaert:2005in}. These equations easily generalise to any dimension and can be
obtained \cite{Bekaert:2005in} from the double-scaling limit (zero-mass and infinite-spin limit, with their product being fixed) of Khan and Ramond \cite{Khan:2004nj}.
The Segal-like action of Schuster and Toro \cite{Schuster:2014hca} (see also \cite{Bekaert:2015qkt} for fermions) that reproduces the equations of motion from \cite{Bekaert:2005in} was used to couple continuous-spin fields with  spinor fields \cite{Kundu:2025mzm}, point-like particles \cite{Schuster:2023xqa} or external currents \cite{Schuster:2023xqa,Schuster:2023jgc,Kundu:2025fsd} (there are, however, some issues with current exchanges \cite{Bekaert:2017xin} and minimal couplings \cite{Rivelles:2016rwo}).
Another formulation are the Fronsdal-like actions by Metsaev \cite{Metsaev:2016lhs,Metsaev:2017ytk,Metsaev:2019opn} which apply as well to Minkowski and (anti) de Sitter\footnote{Note that we will stick here to the continuous-spin massless representations of the Poincar\'e group found by Wigner, and we will not consider their generalizations for the isometry groups of (A)dS spacetimes.} spacetimes. Several BRST approaches have been pushed forward \cite{Bengtsson:2013vra,Alkalaev:2017hvj,Metsaev:2018lth,Buchbinder:2018yoo,Alkalaev:2018bqe,Burdik:2019tzg,Buchbinder:2020nxn,Buchbinder:2024jpt}.
One of the foremost formulation is the light-cone gauge approach by Metsaev \cite{Metsaev:2017cuz,Metsaev:2017myp,Metsaev:2019opn,Metsaev:2021zdg,Metsaev:2025qkr,Metsaev:2025nbm} which was instrumental in his classification of cubic vertices \cite{Metsaev:2017cuz,Metsaev:2018moa,Metsaev:2025qkr,Metsaev:2025wcv}.
Finally, other formulations have also been advocated, using e.g. the frame-like formalism \cite{Khabarov:2017lth}, the method of coadjoint orbits \cite{Gracia-Bondia:2017fai,Gracia-Bondia:2018hrq,Joung:2024akb}, or a pair of commuting variables made of a vector and a spinor \cite{Buchbinder:2023qog}. 

\newpage

Aside from this list of field-theoretic formulations, an alternative strategy is to consider directly on-shell scattering amplitudes, for which one only needs asymptotic states.
Scattering amplitudes of continuous-spin particles have been advocated by various authors over the years to possibly be well-behaved (cf. the early work \cite{Iverson:1971hq} or the more recent works \cite{Schuster:2013pxj,Bellazzini:2022wzv,Schuster:2023xqa,Schuster:2023jgc,Bellazzini:2024dco}), i.e. obeying---at least some of---the standard conditions on the S-matrix (such as Poincar\'e invariance, analyticity or unitarity) as well as the ``helicity correspondence'', i.e. recovering the results of ordinary massless particles in the limit $E\gg \mu$ where the energy scale $E$ is much higher than the continuous-spin scale $\mu$ (the norm of the Pauli--Lubanski vector). To date, the most systematic exploration of such on-shell scattering amplitudes is the work \cite{Bellazzini:2024dco}, the results of which confirm the expectation that continuous-spin particles might provide a new infrared deformation of ultraviolet theories, where infinitely-many degrees of freedom appear at low energy ($E\sim \mu$) but effectively decouple at high energy ($E\gg\mu$). 
In fact, the infrared physics of gravity and gauge theories provides another motivation for studying the scattering of continuous-spin particles, as will now be explained.

Originally, the generalization to continuous-spin particles \cite{Schuster:2013pxj} of Weinberg's soft theorems \cite{Weinberg:1964ew} was one of the seminal results which prompted the surge of interest in these exotic massless representations because their soft limit does \textit{not} imply any conservation law and, consequently, they bypass the usual no-go theorems against massless particles with helicities higher than two (see e.g. the review \cite{Bekaert:2010hw} on the latter). The status of the soft limit of scattering amplitudes of continuous-spin particles remains a crucial issue for assessing their viability as infrared deformations of usual massless particles.

Our own motivation for studying their scattering is grounded in the contemporary perspective on Weinberg's soft theorems as Ward identities of asymptotic symmetries and, more specifically, in the modern insight \cite{Strominger:2013jfa,He:2014laa} that the Bondi--Metzner--Sachs (BMS) group \cite{bondi_gravitational_1962,Sachs:1962wk} should be a symmetry of any scattering problem involving gravity. 
If BMS transformations are taken as genuine symmetries of the S-matrix, then it becomes natural to organise the asymptotic states into unitary irreducible representations of the BMS group, i.e. ``BMS particles''. Accordingly, the representation theory of the BMS group (see \cite{Girardello:1974sq,McCarthy_75} and refs therein) was revisited, bringing in modern physical insights \cite{Bekaert:2024uuy,Bekaert:2025kjb}. Endowed with this reorganization of asymptotic states, a natural program of research would be a version of the S-matrix bootstrap where Poincar\'e symmetry gets replaced by BMS symmetry, i.e. a ``(BM)S-matrix bootstrap''. To free oneselves of any Poincar\'e prejudice, one might thus consider the scattering of \textit{generic} BMS particles.
An important feature of such BMS particles is that they typically form infinite-dimensional BMS multiplets of Poincar\'e particles (in the sense that a generic UIR of the BMS group decomposes into an infinite collection of UIRs of the Poincar\'e group).
In particular, a massless BMS multiplet generically includes an infinite tower of continuous-spin particles with a continuum of values of the spin scale \cite{McCarthy_73-III,Bekaert:2025kjb}. 
Consequently, a prerequisite of a systematic survey of the scattering properties of BMS particles is the study of continuous-spin particle scattering amplitudes.

In the present paper, we would like to propose a new formulation of continuous-spin particles which makes use of a  complex two-component spinor to build asymptotic states and allows to write down on-shell amplitudes in a straightforward way. 
Using a pair of commuting spinor variables for describing massive particles is by-now standard in the amplitude literature (see e.g. \cite{Dittmaier:1998nn,Conde:2016vxs,Conde:2016izb,Arkani-Hamed:2017jhn}). Furthermore, pairs of two-component spinor variables\footnote{Although spinors variables come handy to address supersymmetry, supermultiplets of continuous-spin fields \cite{Brink:2002zx,Buchbinder:2019esz,Buchbinder:2019sie,Buchbinder:2019iwi,Najafizadeh:2019mun,Najafizadeh:2021dsm,Buchbinder:2022msd,Buchbinder:2025ztn} will not be considered here.} were used previously for describing continuous-spin particles (see e.g. \cite{Buchbinder:2018soq,Buchbinder:2019iwi,Buchbinder:2019sie,Bellazzini:2024dco}) but the formulation presented here is distinct in that it eventually makes of use of a single spinor variable. 

The present formulation was found by applying Wigner's method of induced representations and proves useful for unifying the description of massless particles (the usual helicity representations and the continuous-spin ones). One starts with a wavefunction on the Lorentz group, thereby depending on a pair of two-component spinors that parameterizes an arbitrary element of $GL(2,\mathbb{C})\supset SL(2,\mathbb{C})$. In the case of massless representations (helicity or continuous-spin), the equivariance of wavefunctions under the little group $ISO(2)\subset SO(3,1)$ implies that they can be reduced, without loss of generality, to functions of a single two-component spinor. Basically, the main technical simplification of continuous-spin wavefunctions achieved by the formulation proposed here is that they can be described by an unconstrained function $\Phi(\lambda)$ of a \textit{single} two-component spinor $\lambda_\alpha$. What changes, as compared to usual massless particles, is the action of Lorentz generators, which now reads
$$
    P_{\alpha\dot\alpha} = \lambda_\alpha \bar\lambda_{\dot\alpha}\,, \qquad
    J_{\alpha\beta} = 2i\,\lambda_{(\alpha}\epsilon_{\beta)\gamma}\,
    \frac{\partial}{\partial\lambda_\gamma}
    - \frac{\mu}{\langle\lambda\xi\rangle^2}\,\xi_\alpha \xi_\beta\,,
$$
where $\xi_\alpha$ is an auxiliary ``reference'' spinor. The standard Lorentz generator gets deformed by a simple correction which is proportional to the continuous-spin scale $\mu$. The helicity content of a wave function $\Phi(\lambda)$ corresponds to its expansion in homogeneous components.

This technical simplification allows us to easily write down Lorentz-invariant on-shell amplitudes of massless particles including continuous-spin ones. In this sense, the present formulation can be thought as an extension of the spinor-helicity formalism that allows to include continuous-spin particles. We use this powerful tool to reconsider the problem of building on-shell amplitudes including continuous-spin particles and some new results are obtained. 

More precisely, we investigate three-point (and higher-point) amplitudes and compared our findings with the most recent results available in the literature, on cubic vertices \cite{Metsaev:2025wcv} and on scattering amplitudes \cite{Bellazzini:2024dco}, finding agreement with those works.
In particular, as was found in \cite{Bellazzini:2024dco} the on-shell amplitudes reduce to the amplitudes of usual massless particles with the corresponding helicities, up to a kinematical 
prefactor which has the form of an exponent of certain simple functions of momentum spinors, weighted by the continuous-spin scales of the corresponding external particles. These dressings by a prefactor have been first observed in three-point interactions \cite{Metsaev:2017cuz,Metsaev:2018moa}. We also show that the exponential prefactor emerges in the infinite-spin zero-mass limit of usual amplitudes.

Furthermore, new three-point amplitudes are found which circumvent some of the kinematical obstructions observed in \cite{Bellazzini:2024dco} and follow the idea of distributional solutions of \cite{Metsaev:2017cuz,Metsaev:2018moa,Metsaev:2025wcv}. The two special cases are: (1) three massless particles with at least two of continuous-spin type (in case there is just one continuous-spin particle, no solution seem to exist); (2) one continuous-spin particle and two massive particles with equal mass. Both cases have some kinematical degeneracy: all momenta are collinear for (1) and the massive momenta are collinear to the massless one for (2). It is easy to see in both cases there are nontrivial three-point amplitudes, but Lorentz symmetry imposes further kinematical constraints on the scattering. Some higher-point amplitudes are also proposed, which are inspired by the recent discovery of nonvanishing single-minus amplitudes in \cite{Guevara:2026qzd} and by their higher-spin counterparts \cite{Ponomarev:2022atv,Serrani:2026azw}.

The paper is organized as follows. In section \ref{sect2}, we briefly review Wigner's method of induced representations for constructing unitary and irreducible representations of the Poincar\'e group and apply it to the continuous-spin case in order to push forward a realization in terms of wavefunctions depending on a single two-component spinor.
The mathematical equivalence of this representation with the one in terms of a pair of such spinors is discussed, as well as the decomposition in helicity eigenstates.
Section \ref{sect4} is devoted to the analysis of on-shell amplitudes involving at least one continuous-spin particle as external leg. Their general form is discussed in section \ref{Generalform} while explicit examples are provided in section \ref{Explicitexamples}. We conclude in section \ref{concl} with a brief  summary of our findings as well as some speculations on future developments. Some technical results have been moved to the appendix.

%%%%%%%%%%%%%%%%%%%%%%%%%%%%%%%%%%%%%%%%
\section{Continuous-spin representation}\label{sect2}
%%%%%%%%%%%%%%%%%%%%%%%%%%%%%%%%%%%%%%%%
Unitary and irreducible representations of the Poincar\'e group
are famously exhausted by induced representations \cite{Wigner:1939cj},
where the inducing subgroup
is taken to be the stabiliser of a momentum orbit, aka ``little group'' denoted $H$, i.e. $H=SO(3)$ for massive particles and $H=ISO(2)$ for massless particles.

Such induced representations \cite{Wigner:1939cj,MacKey} are defined as functions
on the (double cover of the) Lorentz group, $SL(2,\mathbb{C})$,
valued in a given unitary and irreducible representation $V$
of the little group $H\subset SL(2,\mathbb{C})$ with action $D_V: H \longrightarrow U(V)$,
and equivariant under the right action of the latter,
\begin{equation}
    \mathcal{F}_V
    := \big\{F \in \mathscr{C}^\infty\big(SL(2,\mathbb{C}),V\big)
    \mid F(\Lambda h) = D_V(h^\inv)F(\Lambda)\,,\ 
    \forall \Lambda \in SL(2,\mathbb{C}),\,h \in H\big\}\,.
\end{equation}
In more mundane terms, the elements of the space $\mathcal{F}_V$ are in one-to-one correspondence with usual on-shell fields, i.e. functions over the momentum orbit $\mathcal{O}_p\cong SL(2,\mathbb{C})/H$ taking values in $V$.
Here we can take advantage of the fact that elements
of $SL(2,\mathbb{C})$ can be parameterised by a pair
of \emph{linearly independent} two-component spinors,
that we shall denote by $\lambda_\alpha$ and $\rho_\alpha$
with $\alpha=1,2$, simply by considering them to be the columns
of a $GL(2,\mathbb{C})$ matrix,
\begin{equation}\label{matrixformed}
    GL(2,\mathbb{C}) \ni \big(\lambda_\alpha\,\ \rho_\alpha\big)
    \equiv \begin{pmatrix}
        \lambda_1 & \rho_1 \\
        \lambda_2 & \rho_2
    \end{pmatrix}\,,
\end{equation}
which we can restrict to $SL(2,\mathbb{C})$ upon requiring
\begin{equation}
    \langle\lambda\rho\rangle
    :=\epsilon^{\alpha\beta}\lambda_\alpha\rho_\beta=1
    \qquad\Longrightarrow\qquad 
    (\lambda_\alpha\,\ \rho_\alpha) \in SL(2,\mathbb{C})\,,
\end{equation}
where $\epsilon^{\alpha\beta}=-\epsilon^{\beta\alpha}$
is the $SL(2,\mathbb{C})$-invariant tensor (with $\epsilon^{12}=1$).%
\footnote{In this paper, we use the convention that
spinor indices are raised and lowed according to,
$\epsilon^{\alpha\beta} \upsilon_\beta=\upsilon^\alpha$
and $\upsilon^\alpha\epsilon_{\alpha\beta}=\upsilon_\beta$.
Accordingly, one has $\epsilon^{\alpha\gamma}\epsilon_{\beta\gamma}=\delta^\alpha_\beta$. On top of the angle bracket
denoting the contraction of \emph{undotted spinors}, we will also use the square bracket notation $[\bar\lambda\bar\xi]:=\epsilon^{\dot\alpha\dot\beta} \bar\lambda_{\dot\alpha}\bar\xi_{\bar\beta}$ to denote the contraction of \emph{dotted spinors} (in particular, note that $\overline{\langle\lambda\xi\rangle}=[\bar\lambda\bar\xi]$ for commuting spinors).}
Another possibility is to consider functions
of two linearly independent spinors, which is simply requiring
$\langle\lambda\rho\rangle\neq0$ without necessarily fixing
this angle-bracket to $1$, but instead demanding that
the functions of pairs of spinors be homogeneous
under rescaling of $\lambda$ and $\rho$ by a non-zero complex number.
In other words, functions of $(\lambda,\rho)$
with $\langle\lambda\rho\rangle\neq0$ are functions
on $GL(2,\mathbb{C})$, which upon requiring them to verify
\begin{equation}
    F(t\,\lambda,t\,\rho,\bar t\,\bar\lambda, \bar t\,\bar\rho)
    = |t|^w\,F(\lambda,\rho,\bar\lambda,\bar\rho)\,,
    \qquad \forall t \in \mathbb{C}^\times\,,
\end{equation}
for some $w \in \mathbb{R}$, descend to functions
on $GL(2,\mathbb{C})/\mathbb{C}^\times \cong SL(2,\mathbb{C})$.
In the rest of the paper, we will opt for the first option,
which is to say that we will assume $\langle\lambda\rho\rangle=1$. Moreover, we work with functions of complex variables 
throughout and, for conciseness, omit the complex conjugates 
from the notation, even when the functions are not holomorphic.

The action of the Poincar\'e group on such functions
is then given by
\begin{equation}
    \big[U(\Lambda,a)F\big](\lambda,\rho)
    = e^{ia^\mu p_\mu(\lambda,\rho)}\,
    F(\Lambda^\inv\lambda,\Lambda^\inv\rho)\,,
\end{equation}
where $\Lambda\in SO(3,1)$ is a Lorentz transformation,
$a\in\mathbb{R}^4$ is a translation and $p_\mu(\lambda,\rho)$
is the momentum associated with the pair of spinors $(\lambda,\rho)$. 
More specifically, 
\begin{equation}
    p_\mu(\lambda,\rho) = \tfrac12\bar\sigma_\mu^{\alpha\dot\alpha}
    \lambda^I_\alpha\tilde\lambda_{I\dot\alpha}\,,
    \qquad\text{with}\qquad
    \lambda^{I=1}_\alpha=\lambda_\alpha\,,
    \quad 
    \lambda^{I=2}_\alpha = m\,\rho_\alpha\,,
    \quad\text{and}\quad
    \tilde\lambda_{I\dot\alpha} = \bar\lambda^I_\alpha\,,
\end{equation}
with $m$ is the mass of the particle of interest,
and $\sigma_\mu=(\mathbf{1},\vec\sigma)$ are
the usual four matrices consisting of the identity
and the three Pauli matrices with components $\sigma_{\mu\,\alpha\dot\alpha}$, and $\bar\sigma_\mu^{\alpha\dot\alpha}:=\epsilon^{\alpha\beta}\epsilon^{\dot\alpha\dot\beta}\sigma_{\mu\,\beta\dot\beta}$.\footnote{The norm of four-vectors in Minkowski spacetime with mostly minus signature is obtained by using $\eta^{\mu\nu}\sigma_\mu^{\alpha\dot\alpha}\sigma_\nu^{\beta\dot\beta}=2\epsilon^{\alpha\beta}\epsilon^{\dot\alpha\dot\beta}$ as $p^2=\det(p_{\alpha\dot\alpha})\equiv \tfrac12\,p_{\alpha\dot\alpha}p^{\alpha\dot\alpha}$. Consequently, one verifies that $p^2=m^2$ with the normalisation $\lambda^I\lambda^J=m\,\epsilon^{IJ}$ used in this paper.}
For massive fields, this reproduces the standard spinor-helicity
representations wherein one uses a pair of two-component spinors to parameterise
massive momenta, and consider this pair as an $SU(2)$-doublet 
$\lambda_\alpha^I$ with $I=1,2$ an index for the fundamental
representation of $SU(2)$
\cite{Dittmaier:1998nn, Conde:2016izb, Arkani-Hamed:2017jhn}.%
\footnote{Note that in the usual spinor-helicity formalism for massive particles, the \emph{complex} momentum is recovered from $p_{\alpha\dot\alpha} = \lambda^I_\alpha\tilde\lambda_{I\dot\alpha}$ with the tilde spinors $\tilde\lambda_{I\dot\alpha}$ considered as independent from the untilde ones $\lambda^I_\alpha$. Since here we are working with representations of the real Poincar\'e group and not its complexification, we have already imposed the reality conditions $\tilde\lambda_{I\dot\alpha}=\bar\lambda^I_\alpha$ on the previous expression, thereby ensuring that the momentum is \emph{real}. We have also introduced the mass explicitly in the definition of $\lambda^2_\alpha$ in order to both have a smooth massless limit and make sure to recover the mass squared while having fixed $\langle\lambda\rho\rangle=1$.}
For massless fields, only one spinor is necessary to parameterise
massless momenta, so this representation may seem `too large'.
As it turns out, the $ISO(2)$-equivariance property implies
that this additional spinor can be gauged-away. 

Recall that we have two types of massless particles,
depending on the type of representation of $ISO(2) \equiv SO(2) \ltimes \mathbb{R}^2$ that we choose.
\begin{itemize}
\item {\bf Helicity.} The Euclidean group in two dimensions,
$ISO(2)$, admits a family
of one-dimensional unitary irreducible representations wherein its $\mathbb{R}^2$
factor is represented trivially, so that it effectively
reduces to the unitary irreducible representations of its $SO(2) \cong U(1)$ piece.
The latter acts by a phase depending on an integer
$h \in \tfrac12\,\mathbb{Z}$, the helicity, i.e.
\begin{equation}\label{eq:h}
    D_h(\varphi,z) = e^{ih\varphi}\,,
\end{equation}
with $\varphi \in [0,2\pi)$ and $z \in \mathbb{C}$ are
the parameters of, respectively, the $SO(2) \cong U(1)$
and $\mathbb{R}^2 \cong \mathbb{C}$ subgroups.

\item {\bf Continuous-spin.} The other family of unitary irreducible representations
of the Euclidean group is faithful, i.e. the two-dimensional
translations are now represented non-trivially,
and the representation is infinite-dimensional. Indeed, the Euclidean group  can be realized
on the space of square-integrable functions on the circle $S^1$,
with the action of $ISO(2)$ given by
\begin{equation}\label{eq:mu}
    \big[D(\varphi, z)f\big](\theta)
    = e^{i\mu (z e^{-i\theta}+\bar z e^{i\theta})/2}
    f(\theta+\varphi)\,,
    \qquad f \in L^2(S^1)\,,
    \qquad \mu \in \mathbb{R}_+\,,
\end{equation}
i.e. the translation part of the $ISO(2)$ acts by multiplication
with a phase, of argument proportional to
$(z e^{-i\theta}+\bar z e^{i\theta})/2 \equiv \Re(ze^{-i\theta})$.
The parameter $\mu$ corresponds to the continuous-spin scale, when using this irreducible representation of $ISO(2)$ to induce one of the Poincar\'e group $ISO(3,1)$.
\end{itemize}

Using the embedding of $ISO(2)$ into $SL(2,\mathbb{C})$
as the upper-triangular matrices of the form
\begin{equation}
    ISO(2) \ni (\varphi,z) \quad\longrightarrow\quad
    \begin{pmatrix}
        e^{-i\varphi/2} & e^{i\varphi/2}z \\
        0 & e^{i\varphi/2}
    \end{pmatrix}
    \in SL(2,\mathbb{C})\,,
\end{equation}
the equivariance property for functions of $SL(2,\mathbb{C})$,
understood as functions of a pair of spinors, valued
in either types of representations discussed above, reads respectively as
\begin{subequations}
\label{eq:equivariance_iso(2)}
\begin{align}
    F\big(e^{-i\varphi/2}\lambda,
    e^{i\varphi/2}(\rho+z\lambda)\big)
    & = e^{ih\varphi} F(\lambda,\rho)\,,
    && \text{[Helicity]} \\
    F\big(e^{-i\varphi/2}\lambda,
    e^{i\varphi/2}(\rho+z\lambda),\theta\big)
    &= e^{-i\mu\Re(ze^{-i(\theta-\varphi)})}
        F(\lambda,\rho,\theta-\varphi)\,,
    && \text{[Continuous-spin]}
\end{align}
\end{subequations}
where $h$ is the helicity and $\mu$ is the continuous-spin scale. Their infinitesimal version leads to the following conditions,
\begin{subequations}
\begin{align}
    \label{eq:translation}
    0 & = \big(\lambda_\alpha\tfrac{\partial}{\partial \rho_\alpha}
    + \tfrac{i}2 \mu e^{-i\theta}\big)F\,,\\
    \label{eq:translation_cplx}
    0 & = \big(\bar\lambda_{\dot\alpha}\tfrac{\partial}{\partial \bar\rho_{\dot\alpha}}
    + \tfrac{i}2 \mu e^{+i\theta}\big)F\,,\\
        \label{eq:rotation_helicity}
    0 & = \big(\lambda_\alpha\tfrac{\partial}{\partial\lambda_\alpha}
    -\bar\lambda_{\dot\alpha}\tfrac{\partial}{\partial\bar\lambda_{\dot\alpha}}
    -\rho_\alpha\tfrac{\partial}{\partial\rho_\alpha}
    +\bar\rho_{\dot\alpha}\tfrac{\partial}{\partial\bar\rho_{\dot\alpha}}+2h\big)F\,, && \text{[Helicity]}\\
    0 & = \big(\lambda_\alpha\tfrac{\partial}{\partial\lambda_\alpha}
    -\bar\lambda_{\dot\alpha}\tfrac{\partial}{\partial\bar\lambda_{\dot\alpha}}
    -\rho_\alpha\tfrac{\partial}{\partial\rho_\alpha}
    +\bar\rho_{\dot\alpha}\tfrac{\partial}{\partial\bar\rho_{\dot\alpha}}
    +2i\tfrac{\partial}{\partial\theta}\big)F\,, && \text{[Continuous-spin]}
    \end{align}
\end{subequations}
where the first two equations come from the $ISO(2)$-translations
(including the helicity case for $\mu=0$), and the last two equations
from the $SO(2)$ rotations in the helicity and continuous-spin cases, respectively.

For helicity-type particles, conditions
\eqref{eq:translation} and \eqref{eq:translation_cplx}
reduce to $ISO(2)$-translation \emph{invariance}, which allows
one to get rid of the $\rho$ dependency altogether.
Indeed, that $F(\lambda,\rho)$ is annihilated by
$\lambda_\alpha\tfrac{\partial}{\partial\rho_\alpha}$
and its complex conjugate implies that it depends
on $\rho$ only through the combination $\langle\lambda\,\rho\rangle$,
which is already fixed to $1$ by assumption. The condition
\eqref{eq:rotation_helicity} then fixes the homogeneity
in $\lambda$ and $\bar\lambda$ with respect to the helicity,
thereby leading to
\begin{equation}\label{eq:massless_helicity}
    F(\lambda,\bar\lambda) = \left\{
    \begin{aligned}
        \lambda_{\alpha_1} \dots \lambda_{\alpha_{2h}}\,
        F^{\alpha_1 \dots \alpha_{2h}}(\lambda_\beta\bar\lambda_{\dot\beta})\,,
        && h \leqslant 0\,,\\
        \bar\lambda_{\dot\alpha_1} \dots \bar\lambda_{\dot\alpha_{2h}}\,
        F^{\dot\alpha_1 \dots \dot\alpha_{2h}}(\lambda_\beta\bar\lambda_{\dot\beta})\,,
        && h \geqslant 0\,.
    \end{aligned}
    \right.
\end{equation}

In order to solve the constraints \eqref{eq:translation}
and \eqref{eq:translation_cplx} for $\mu\neq0$, let us introduce
a fixed, reference, two-component spinor $\xi \in \mathbb{C}^2\backslash\{(0,0)\}$
and work in the patch $\langle\lambda\xi\rangle\neq0$.
Having done so, we can re-express the second spinor $\rho$ as
\begin{equation}\label{eq:def_u}
    \rho = \tfrac1{\langle\lambda\xi\rangle}\xi - u\,\lambda\,,
    \qquad 
    u := \frac{\langle\xi\,\rho\rangle}{\langle\lambda\,\xi\rangle}\,,
\end{equation}
i.e. we express it in the basis of $\mathbb{C}^2$
defined by $\lambda$ and $\xi$.%
\footnote{Note that under the action of $SL(2,\mathbb{C})$
on $\lambda$ and $\rho$, the complex number $u$ transforms
through the associated M\"obius transformation, $u\big(\Lambda^\inv\lambda,\Lambda^\inv\rho\big)
    = \frac{\langle\xi\,\Lambda^\inv\rho\rangle}{\langle\Lambda^\inv\lambda\,\xi\rangle} = \frac{au+b}{cu+d}
    \equiv \mathcal{M}(\Lambda,u)$, with $\Lambda=\begin{pmatrix}
       a & b \\
        c & d
    \end{pmatrix}$.}
In terms of the variables $(\lambda,u,\theta)$,
the $ISO(2)$-translation equivariance reads
\begin{equation}
    \big(\tfrac{\partial}{\partial u}
        - \tfrac{i}{2}\mu e^{-i\theta}\big)F = 0
    = \big(\tfrac{\partial}{\partial \bar u}
        - \tfrac{i}{2}\mu e^{i\theta}\big)F
    \qquad\Longrightarrow\qquad
    F(\lambda,u,\theta) = e^{i\mu\Re(ue^{-i\theta})}
    G(\lambda,\theta)\,,
\end{equation}
for some other function $G$, independent of $u$.
Equivariance under $SO(2) \subset ISO(2)$ boils down to
\begin{equation}
    \big(\lambda_\alpha\tfrac{\partial}{\partial\lambda_\alpha}
    -\bar\lambda_{\dot\alpha}\tfrac{\partial}{\partial\bar\lambda_{\dot\alpha}}
    +2i\tfrac{\partial}{\partial\theta}\big)G = 0\,,
\end{equation}
i.e. it fixes the $\theta$ dependency of $G$,
so that we end up with
\begin{equation}\label{eq:u}
    F(\lambda,u,\theta)
    = e^{i\mu\Re(ue^{-i\theta})}\Phi(e^{i\theta/2}\lambda)\,.
\end{equation}
Note that due to the phase of $\theta/2$ appearing
in front of the dependency on $\lambda$, the above function $F$
further verifies
\begin{equation}
    F(\lambda,u,\theta+2\pi) = F(-\lambda,u,\theta)\,,
\end{equation}
which signals that we should distinguish between \emph{even}
and \emph{odd} functions of $\lambda$, corresponding respectively
to `bosonic' and `fermionic' continuous-spin fields. 

%======================================%
\paragraph{Equivalence of realizations:}
%======================================%
We can think of this relation \eqref{eq:u} as defining an isomorphism $\mathcal{L}_{(\mu)}:\Phi\mapsto F$
between the space of functions $F(\lambda,u,\theta)$ obeying the equivariance condition \eqref{eq:equivariance_iso(2)}%
---expressed in the $(\lambda,u,\theta)$ variables---%
and the space of functions $\Phi(\lambda)$ of a single spinor,
\begin{equation}\label{eq:isomorphism}
    \mathcal{L}_{(\mu)}[\Phi](\lambda,u,\theta)
    := e^{i\mu\Re(ue^{-i\theta})}\,\Phi(e^{i\theta/2}\lambda)\,,
\end{equation}
whose inverse $\mathcal{L}_{(\mu)}^\inv:F\mapsto\Phi$ is simply given by the evaluation at $u=0=\theta$,%
\footnote{Note that setting $u=0$ amounts to fixing $\rho$ is collinear with $\xi$ (i.e.
$\rho \propto \xi$) due to its definition \eqref{eq:def_u}.}
\begin{equation}
    \mathcal{L}_{(\mu)}^\inv[F](\lambda) \equiv F(\lambda,0,0)\,.
\end{equation}
Using this isomorphism to transfer the action of the Poincar\'e group
on functions of a single spinor, one finds
\begin{equation}
\label{eq:CSPPoincare}
    \big[U^{\mathcal{L}_{(\mu)}}(\Lambda,a)\Phi\big](\lambda)
    = e^{ia^\mu p_\mu(\lambda)}\,\exp\Big(i\mu\,
    \Re\Big[\frac{\langle\xi\Lambda^\inv\xi\rangle}
    {\langle\xi\lambda\rangle\langle\xi
    \Lambda^\inv\lambda\rangle}\Big]\Big)\,
    \Phi(\Lambda^\inv\lambda)\,.
\end{equation}
One can check that this \emph{does define} a representation
of the Poincar\'e group as a consequence of
the Schouten identity for two-component spinors
(needed in order to ensure that the $\xi$-dependent
phase factor recombines correctly), and thus we find that continous-spin wavefunctions can be realized as arbitrary functions of a single spinor $\lambda$ transforming under Poincaré as~\eqref{eq:CSPPoincare}.
The infinitesimal form of this action reads
\begin{subequations}
\begin{align}
    P_{\alpha\dot\alpha} & = \lambda_\alpha \bar\lambda_{\dot\alpha}\,,\\
    J_{\alpha\beta} & = 2i\,\lambda_{(\alpha}\epsilon_{\beta)\gamma}\,
    \frac{\partial}{\partial\lambda_\gamma}
    - \frac{\mu}{\langle\lambda\xi\rangle^2}\xi_\alpha \xi_\beta\,,\\
    \bar J_{\dot\alpha\dot\beta} & = 2i\,\bar\lambda_{(\dot\alpha}\epsilon_{\dot\beta)\dot\gamma}\,
    \frac{\partial}{\partial\bar\lambda_{\dot\gamma}}
    - \frac{\mu}{[\bar\lambda\bar\xi]^2}
    \bar\xi_{\dot\alpha} \bar\xi_{\dot\beta}\,.
\end{align}
\end{subequations}
where the round bracket around the indices denotes their symmetrization with weight one, i.e. $S_{(\alpha\beta)}=\frac12(S_{\alpha\beta}+S_{\beta\alpha})$. The $\mu$-deformed Lorentz action here-above is, perhaps, the most important formula in the paper. For example, amplitudes of the next section are just solutions to the Poincar\'e  invariance. 

Note that the generators $J_{\alpha\beta}$ and $\bar J_{\dot\alpha\dot\beta}$ do obey the same $\mathfrak{sl}(2,\mathbb{C})$ commutation relations as their first terms, which can be checked directly and the second order pole of the second term is fixed by the closure of the algebra.
Another consistency check is performed by computing the eigenvalue of the quartic Casimir operator of the Poincar\'e operator. The square of the Pauli--Lubanski vector,
\begin{equation}\label{PLvector}
W_{\alpha\dot\alpha}:=\frac{i}{2}\left(J_{\alpha\beta}P^\beta{}_{\dot\alpha}-\bar J_{\dot\alpha\dot\beta}P_\alpha{}^{\dot\beta}\right)=-{\frac{1}{2}}P_{\alpha\dot\alpha}\left(\lambda_\alpha\tfrac{\partial}{\partial\lambda_\alpha}
    -\bar\lambda_{\dot\alpha}\tfrac{\partial}{\partial\bar\lambda_{\dot\alpha}}\right)
+\frac{i\mu}{2}\left(\frac{\xi_\alpha\bar\lambda_{\dot\alpha}}{\langle\lambda\xi\rangle} -\frac{\lambda_{\alpha}\bar\xi_{\dot\alpha}}{[\bar\lambda\bar\xi]}\right)\,,
\end{equation}
is equal to
\begin{equation}
W^2:=\frac12 W_{\alpha\dot\alpha}W^{\alpha\dot\alpha}=-\frac{\mu^2}{4},
\end{equation}
as it should in our conventions.\footnote{Note that, in our conventions, the continuous-spin scale $\mu$ is twice its usual value in the literature.} Also note that for $\mu=0$, the equation \eqref{PLvector} reproduces the relation $W_{\alpha\dot\alpha}=h\, P_{\alpha\dot\alpha}$ of helicity-type particles.

In order to understand the relation between the above
isomorphism between $ISO(2)$-equivariant functions $F$ on $SL(2,\mathbb{C})$, valued in functions on the circle, and functions $\Phi$ of a single spinor
$\lambda \in \mathbb{C}^2\backslash\{(0,0)\}$, let us observe
that a choice of a non-vanishing spinor $\xi$ such that 
$\langle\lambda\,\xi\rangle\neq0$ allows one to write down
the decomposition
\begin{equation}\label{4variables}
    \mathbb{C}^2\backslash\{(0,0)\} 
        \cong
        S^2 \times \mathbb{R}_+ \times S^1\,,
\end{equation}
where the last two factors merely correspond
to decomposing a non-zero complex number into its modulus
and a phase. Concretely, $\xi$ can be completed into a basis
of $\mathbb{C}^2\backslash\{(0,0)\}$ by $\xi_\perp$
such that $\langle\xi\,\xi_\perp\rangle=1$,
so that the spinor $\lambda$ can be written as
\begin{equation}
    \lambda = \langle\lambda\,\xi_\perp\rangle\,\xi 
    - \langle\lambda\,\xi\rangle\,\xi_\perp
    \equiv \sqrt{\omega}\,e^{i\phi}\,\big(z\,\xi-\xi_\perp\big)\,,
\end{equation}
where we introduced
\begin{equation}
    z := \frac{\langle\lambda\,\xi_\perp\rangle}{\langle\lambda\,\xi\rangle}\,,
    \qquad 
    \sqrt{\omega} := \big\lvert\langle\lambda\,\xi\rangle\big\rvert\,,
    \qquad
    e^{i\phi} := \frac{\langle\lambda\,\xi\rangle}{\lvert\langle\lambda\,\xi\rangle\rvert}\,
\end{equation}
which can respectively be understood as coordinates
on $\mathbb{C}\cup\{\infty\}\cong S^2$, on $\mathbb{R}_+=(0,\infty)$ and on $S^1$. The first two coordinates, 
namely $z$ and $\omega$, correspond to coordinates
on the null momentum orbit, i.e. the following coset diffeomorphic to a cone
\begin{equation}
    SL(2,\mathbb{C})/ISO(2) \cong S^2 \times \mathbb{R}_+\,,
\end{equation}
with $\omega$ the energy and $z$ defining a null direction
(by specifying a point on the celestial sphere). The dependency 
on the remaining phase is usually fixed by requiring wavefunctions
for helicity-type massless particles to be homogeneous
under rescaling of $\lambda$ by a phase or, equivalently,
to be equivariant under the action of the $SO(2)$-piece 
of the little group as in \eqref{eq:equivariance_iso(2)}.
By relaxing this condition, we effectively keep a dependency
on an additional factor $S^1$ as in \eqref{4variables}, which agrees with
the counting of the number of variables a continuous-spin particle
wavefunction depends on: either a complex 2-component spinor (in our equivalent description) or a null 4-momentum together with an angle for the circle (in the traditional description), i.e. four independent real variables in any formulation.

%=========================%
\paragraph{Helicity limit:}
%=========================%
Note that this picture is in accordance with the fact
that the faithful, infinite-dimensional, representation
\eqref{eq:mu} of $ISO(2)$, defining a continuous-spin particle,
admits a contraction to the direct sum of the unfaithful ones 
\eqref{eq:h} for all helicities. Indeed, taking the limit $\mu\to0$
on the representation \eqref{eq:mu}, one can expand functions $f(\theta)$
on $S^1$ in Fourier series, and notice that each mode $e^{in\theta}$ defines
a one-dimensional invariant subspace carrying the representation
\eqref{eq:h} of helicities equal to half of the Fourier mode, i.e. $h=\frac{n}2$.
In other words, a continuous-spin particle can be thought of
as a superposition of massless particles with all possible
helicities, which are mixed-up altogether under the action
of $SL(2,\mathbb{C})$. Allowing a function of a single spinor
without imposing any homogeneity condition can therefore
be thought of as keeping all possible helicities,
necessary to realize the continuous-spin representation.

Note that the expansion of $\Phi(\lambda)$ in homogeneous components, i.e. eigenstates of the $U(1)$ subgroup action, is only indirectly related to the Fourier mode decomposition of $F(\lambda,u,\theta)$ due to the subtle dependence of $\theta$ in the right-hand-side of \eqref{eq:u}.
Indeed, going back to continuous-spin wavefunctions of the form
\begin{equation}\label{eq:reduced}
    F(\lambda,u,\theta)
    = e^{i\mu\Re(u e^{-i\theta})}\,\Phi(e^{i\theta/2}\lambda)\,,
\end{equation}
they can be decomposed into helicity eigenstates,
which are simply its Fourier modes. Expanding $\Phi$
as a Fourier series,
\begin{equation}
    \Phi(e^{i\theta/2}\lambda) 
    = \sum_{n \in \mathbb{Z}} e^{in\theta}\,\Phi_n(\lambda)\,,
\end{equation}
and using the generating function for Bessel functions
of the first kind, 
\begin{equation}
\label{eq:Bessel}
    e^{\frac{x}{2}(t-1/t)} 
        = \sum_{n \in \mathbb{Z}} J_n(x)\,t^n\,, 
\end{equation}
to expand the exponential factor,
\begin{equation}
    \exp\big(i\mu\Re\big[ue^{-i\theta}\big]\big)
    = \exp\Big[\tfrac{\mu}{2}|u|\big(\tfrac{iu}{|u|}e^{-i\theta}
    -\tfrac{|u|}{iu}e^{i\theta}\big)\Big]
    = \sum_{n \in \mathbb{Z}} \big(\tfrac{iu}{|u|}\big)^n 
    J_n(\mu|u|)\,e^{-in\theta}\,,
\end{equation}
we end up with
\begin{equation}
    F(\lambda,u,\theta) 
    = \sum_{h \in \mathbb{Z}} e^{ih\theta} F_h(\lambda,u)\,,
    \qquad
    F_h(\lambda,u) 
    = \sum_{n \in \mathbb{Z}} \big(\tfrac{iu}{|u|}\big)^n 
    J_n\big(\mu|u|\big)\,\Phi_{h+n}(\lambda)\,.
\end{equation}
Notice that Fourier decomposing $\Phi$ amounts
to decomposing an arbitrary function of $\lambda$
into a sum of \emph{homogeneous} functions (under rescaling
by a phase), of all possible integer degrees.
On a wave function of the form \eqref{eq:reduced},
the Lorentz algebra generators act as
\begin{subequations}
\label{eq:Lorentz_helicity}
\begin{align}
    J_{\alpha\beta} & = \tfrac{2i}{\langle\lambda\xi\rangle^2}\,
    \xi_\alpha \xi_\beta\,\tfrac{\partial}{\partial u} + 2i\,L_{\alpha\beta}\,,
    &&\text{with}&&&
    L_{\alpha\beta} & := \lambda_{(\alpha} \epsilon_{\beta)\gamma}\,
    \tfrac{\partial}{\partial \lambda_\gamma}\,, \\
    \bar J_{\dot\alpha\dot\beta} & = \tfrac{2i}{[\bar\lambda\bar\xi]^2}\,
    \bar\xi_{\dot\alpha} \bar\xi_{\dot\beta}\,\tfrac{\partial}{\partial \bar u} 
    + 2i\,\bar L_{\dot\alpha\dot\beta}\,,
    &&\text{with}&&&
    \bar L_{\dot\alpha\dot\beta} & := \bar\lambda_{(\dot\alpha} \epsilon_{\dot\beta)\dot\gamma}\,
    \tfrac{\partial}{\partial \bar\lambda_{\dot\gamma}}\,,
\end{align}
\end{subequations}
one finds that the Lorentz algebra acts on helicity eigenstates as
\begin{subequations}
\begin{align}
    J_{\alpha\beta} F_h & = 2i\,L_{\alpha\beta} F_h
    -\frac{\mu}{\langle\lambda\xi\rangle^2}
    \xi_\alpha\xi_\beta\,F_{h+1}\,, \label{hel+}\\
    \bar J_{\dot\alpha\dot\beta} F_h 
    & = 2i\,\bar L_{\dot\alpha\dot\beta} F_h
    -\frac{\mu}{[\bar\lambda\bar\xi]^2}
    \bar\xi_{\dot\alpha}\bar\xi_{\dot\beta}\,F_{h-1}\,.\label{hel-}
\end{align}
\end{subequations}
Note that since the $\mu$-dependent part of the Lorentz generators 
changes the helicity by $\pm1$, integer and half-integer helicities
do not mix, which is consistent with the distinction between
`bosonic' and `fermionic' continuous-spin particles
made below \eqref{eq:u}. Note also that it is impossible to restrict the spectrum to all positive/nonnegative helicities (or the opposite) or to all helicities greater or smaller than a given $h$. However, in the complexified setting or in the split signature $J_{\alpha\beta}$ and $\bar J_{\dot\alpha\dot\beta}$ are independent and such truncations seem possible.

When evaluated at $u=0$, the Fourier modes collapse to
\begin{equation}
    F_h\rvert_{u=0} \equiv \Phi_h\,,
\end{equation}
which confirms that the helicity-eigenstates
in the representation in terms of a function
of a single spinor, are nothing but these components
of fixed homogeneity. Moreover, the above formula
shows that eigenstates of different helicities
are mixed-up solely via the $\mu$-dependent part
of the Lorentz generators. This implies that
in the $\mu\to0$ limit, the continuous-spin representation
becomes highly reducible, since each subspace
of fixed helicity is preserved by the action 
of the Poincar\'e group. As these subspaces correspond
to homogeneous functions of $\lambda$,
they are of the form \eqref{eq:massless_helicity},
which to say that in the limit $\mu\to0$,
a continuous-spin representation decomposes
into the direct sum of massless representation
of all possible helicities.

%%%%%%%%%%%%%%%%%%%%
\section{Amplitudes}\label{sect4}
%%%%%%%%%%%%%%%%%%%%

%***********************%
\subsection{General form}\label{Generalform}
%***********************%
Let us consider amplitudes for processes 
involving an arbitrary number of continuous-spin particles
with momentum spinors $\lambda^{(k)}$
and continuous-spin scales $\mu_k$ for $k \in I_{\scriptscriptstyle\sf CSP}$, massive
or massless particles of helicity-type, with momentum spinors
$\chi^{(l)}$ for $l \in I_{\scriptscriptstyle\sf DSP}$, where ``CSP'' stands for the continuous-spin particles and ``DSP'' for the discrete-spin particles,\footnote{Namely, usual particles, i.e. massless particles of helicity-type as well as massive particles. Note that in the latter case, the momentum spinors carry $SU(2)$ indices that we leave implicit in this section, so to lighten the notation.} and a single fixed reference spinor $\xi$. 
Note that we could instead consider one reference spinor $\xi^{(k)}$ for each CSP rather than a single reference spinor $\xi$ shared by all particles. However,  there always exists a reference spinor that is not collinear with any of the spinors $\lambda^{(k)}$ (i.e. $\exists\,\xi\in\mathbb{C}^2$ such that $\langle \xi,\lambda^{(k)}\rangle\neq 0$ for all $k\in I_{\scriptscriptstyle\sf CSP}$) so there is essentially no loss of generality in considering a single reference spinor, as will be done through the paper. 
Let us also stress that, although formulae for individual amplitudes below may appear to depend explicitly on the reference spinor, this is of course a technical artifact: the observables are Lorentz-invariant, and thus the spurious dependence on the reference spinor disappears from the actual final results, such as the total on-shell amplitudes (see e.g. the derivation of \eqref{CSPampli} below).

As usual, we will complexify the representations
and introduce `tilde' spinors $\tilde \lambda^{(k)}$
and $\tilde\chi^{(l)}$, from which we recover 
representations of the real Poincar\'e group by requiring
them to be the complex conjugates of the spinors
$\lambda^{(k)}$ and $\chi^{(l)}$.
On an amplitude describing such scattering, 
the \emph{complexification} of the Poincar\'e group,
\begin{equation}
    ISO(3,1)_{\mathbb{C}}
    \cong \big[SL(2,\mathbb{C}) \times SL(2,\mathbb{C})\big]
    \ltimes \mathbb{C}^4\,,
\end{equation}
acts at the Lie algebra level via
\begin{equation}
    P_{\alpha\dot\alpha} = \sum_{k \in I_{\scriptscriptstyle\sf CSP}}
    \lambda_\alpha^{(k)} \tilde\lambda_{\dot\alpha}^{(k)}
    + \sum_{l \in I_{\scriptscriptstyle\sf DSP}} 
    \chi_{\alpha}^{(l)} \tilde\chi^{(l)}_{\dot\alpha}\,,
\end{equation}
for the complexified translation generators, and 
\begin{subequations}
\begin{align}
    J_{\alpha\beta} & = 2i \,L_{\alpha\beta} 
    - \sum_{k \in I_{\scriptscriptstyle\sf CSP}}
    \frac{\mu_k}{\langle\lambda^{(k)}\xi\rangle^2}\,
    \xi_\alpha\xi_\beta\, ,
    &&&
    L_{\alpha\beta} & = \sum_{k \in I_{\scriptscriptstyle\sf CSP}}
    \lambda^{(k)}_{(\alpha}\epsilon^{\,}_{\beta)\gamma}\,
    \tfrac{\partial}{\partial \lambda^{(k)}_\gamma}
    + \sum_{l \in I_{\scriptscriptstyle\sf DSP}}
    \chi^{(l)}_{(\alpha}\epsilon^{\,}_{\beta)\gamma}\,
    \tfrac{\partial}{\partial \chi^{(l)}_\gamma}\,,\\
    \tilde J_{\dot\alpha\dot\beta} 
    & = 2i\,\tilde L_{\dot\alpha\dot\beta} 
    - \sum_{k \in I_{\scriptscriptstyle\sf CSP}}
    \frac{\mu_k}{[\tilde\lambda^{(k)}\tilde\xi]^2}\,
    \tilde\xi_{\dot\alpha}\tilde\xi_{\dot\beta}\, ,
    &&&
    \tilde L_{\dot\alpha\dot\beta}
    & = \sum_{k \in I_{\scriptscriptstyle\sf CSP}}
    \tilde \lambda^{(k)}_{(\dot\alpha}\epsilon^{\,}_{\dot\beta)\dot\gamma}\,
    \tfrac{\partial}{\partial \tilde\lambda^{(k)}_{\dot\gamma}}
    + \sum_{l \in I_{\scriptscriptstyle\sf DSP}}
    \tilde\chi^{(l)}_{(\dot\alpha}\epsilon^{\,}_{\dot\beta)\dot\gamma}\,
    \tfrac{\partial}{\partial \tilde\chi^{(l)}_{\dot\gamma}}\,,
\end{align}
\end{subequations}
for each one of the two copies of $SL(2,\mathbb{C})$ respectively.
Since the Lorentz generators split into a first-order part,
and a zeroth order (multiplicative) one, the general solution
of Lorentz invariance,
\begin{equation}
\label{eq:LorentzInv}
    J_{\alpha\beta}\,\mathcal{A}_\xi = 0 = \tilde J_{\dot\alpha\dot\beta}\,\mathcal{A}_\xi\,,
\end{equation}
can be written as
\begin{equation}\label{eq:factorisation}
    \mathcal{A}_\xi = e^{-\mathcal{I}_\xi}\,\mathscr{A}\,,
\end{equation}
with 
\begin{subequations}
\label{eq:defining_eq}
\begin{equation}\label{eq:exponent}
    L_{\alpha\beta}\,\mathcal{I}_\xi
    = \frac{i}{2} \sum_{k \in I_{\scriptscriptstyle\sf CSP}}
    \frac{\mu_k }{\langle\lambda^{(k)}\xi\rangle^2}\,
    \xi_\alpha\xi_\beta\,,
    \qquad
    \tilde L_{\dot\alpha\dot\beta}\,\mathcal{I}_\xi
    = \frac{i}{2}\sum_{k \in I_{\scriptscriptstyle\sf CSP}}
    \frac{\mu_k}{[\tilde\lambda^{(k)}\tilde\xi]^2}\,
    \tilde\xi_{\dot\alpha} \tilde\xi_{\dot\beta}\,,
\end{equation}
and
\begin{align}
    L_{\alpha\beta}\mathscr{A} &= 0 = \tilde L_{\dot\alpha\dot\beta}\mathscr{A}\,.
\end{align}
\end{subequations}
Let us note that the translation invariance of the $S$-matrix leads to the momentum conservation realized via $\delta^{(4)}(\sum_i p_i^\mu)$. It can be considered to be a part of $\mathscr{A}$ since the action of the Lorentz generators is canonical for $\mathscr{A}$.

It is clear that any function of the form \eqref{eq:factorisation}
verifying the above conditions \eqref{eq:defining_eq}
is Lorentz-invariant, and conversely, any solution
of the Lorentz invariance constraints can be put in this form,
as long as one can find a solution $\mathcal{I}$ for the first-order partial differential equations\eqref{eq:exponent}.\footnote{We will see below that this is always possible as long as there exists at least two non-collinear spinors that can be formed out of the momentum spinors. The cases in which 
all spinors are collinear will be treated separately later in the text.} This allows us to intertwine
the two representations of the Lorentz algebra given by $J_{\alpha\beta}$
and $L_{\alpha\beta}$, i.e.
\begin{equation}
    J_{\alpha\beta}\,e^{-\mathcal{I}_\xi} = e^{-\mathcal{I}_\xi}\,L_{\alpha\beta}\,,
    \qquad\text{and}\qquad
    \tilde J_{\dot\alpha\dot\beta}\,e^{-\mathcal{I}_\xi} 
    = e^{-\mathcal{I}_\xi}\,\tilde L_{\dot\alpha\dot\beta}\,.
\end{equation}
In other words, since the Lorentz generators can be written
as the sum of the first-order operators $L_{\alpha\beta}$, corresponding to
the usual expression of the Lorentz generators in the spinor-helicity
representations (as if the continuous-spin particles were to be considered
as massless particles of helicity-type), and a multiplicative term,
the amplitude factorises into an exponential factor $e^{-\mathcal{I}}$
and a scattering amplitude term $\mathscr{A}$ for a process
involving the continuous-spin particles \emph{as if they were
massless particles of helicity-type}, except that their helicity%
---and hence the homogeneity in the corresponding momentum spinors%
---is not fixed.

Consequently, the function $\mathscr{A}$ is simply a function
of the angle and square brackets between the various momentum spinors,
and the problem reduces to finding the solutions to equations
\eqref{eq:exponent}. The latter being linear first-order
inhomogeneous partial differential equations, their solutions
take the form of a sum, 
\begin{equation}
    \mathcal{I}_\xi = \mathcal{R}_\xi
    + \mathcal{H}\,,
    \qquad 
    L_{\alpha\beta}\mathcal{H}=0=\tilde L_{\dot\alpha\dot\beta}\mathcal{H}\,,
\end{equation}
with $\mathcal{H}$ a solution of the homogeneous equations
and $\mathcal{R}_\xi$ a particular solution.
The homogeneous solution can be absorbed in $\mathscr{A}$,
so that we should simply find a particular solution.
To construct such a solution, observe that given a 
momentum spinor $\lambda$ of a continuous-spin particle, with associated reference spinor $\xi$,
and any spinor $\psi$ made out of the momentum spinors
(so that it transforms as $L_{\alpha\beta}\psi_\gamma = \psi_{(\alpha}\epsilon_{\beta)\gamma}$),
one has
\begin{equation}
\label{eq:Lratio}
    L_{\alpha\beta} \tfrac{\langle\psi\xi\rangle}
    {\langle\lambda\xi\rangle}
    = -\tfrac{1}{\langle\lambda\xi\rangle^2}
    \xi_{(\alpha}\,\Big[\psi_{\beta)} \langle\lambda\xi\rangle 
    + \lambda_{\beta)} \langle\xi\psi\rangle\Big]
    = \tfrac{\langle\psi\lambda\rangle}{\langle\lambda\xi\rangle^2}\,
    \xi_\alpha \xi_\beta\,,
\end{equation}
upon using Schouten identity in the last step.
For instance, $\psi$ can be the spinor associated to the momentum of another massless particle, be it of continuous-spin or helicity-type. In the latter case, the above ratio is also invariant under `little group scaling', i.e. rescaling the momentum spinors for an helicity-type massless particle by a non-zero complex number, $\chi_\alpha \to t^\inv\chi_\alpha$ and $\tilde\chi_{\dot\alpha} \to t\,\tilde\chi_{\dot\alpha}$ for $t \in \mathbb{C}^\times$, and consequently the exponential of such a ratio automatically satisfies all requirements to be part of the final amplitude. 
One can also consider
\begin{equation}\label{definitionpsi}
   \psi_\alpha = \chi_\alpha^I\,[\tilde\chi_I\tilde\lambda]=p_{\alpha \dot\alpha} \tilde{\lambda}^{\dot\alpha} \,,
\end{equation}
where $\chi_\alpha^I$ is the $SU(2)$-doublet ($I=1,2$ being
the $SU(2)$ indices) of momentum spinors of a massive particle with momentum $p_{\alpha\dot\alpha}$,
which is invariant under the `little group action', i.e. the action of $SU(2)$ on the doublet indices carried by the momentum spinors, and hence can be part of the final amplitude as well.%
\footnote{Note that, if one restricts to constructing such $\psi_A$
as \emph{polynomials} in the various momentum spinors associated
to the particles participating in the process,
then the two elementary solutions discussed above seem to be
the monomials of lowest possible order that can be considered.
Indeed, any monomial of higher order will necessarily be a linear
combination of these, dressed with some monomial in built out
of $SL(2,\mathbb{C})$, and little group, invariant quantities
which are nothing but angle or square bracket of spinor,
with all possible $SU(2)$ indices contracted. Such terms can then
always be absorbed in the homogeneous part of the solution
of \eqref{eq:exponent}.} 

Note that there is a simple reason for \eqref{eq:Lratio} to exist: $\xi_\alpha$ can be expanded in the basis of $\lambda_\alpha$ and $\psi_\alpha$ if $\langle\psi\lambda\rangle\neq0$. Whenever $\langle\psi\lambda\rangle\neq0$, this quantity
being obviously $SL(2,\mathbb{C})$-invariant, one can divide
the ratio on the left-hand side of~\eqref{eq:Lratio} by it to obtain, upon multiplying
by the continuous-spin scale, a particular solution of the equation
\eqref{eq:exponent} for a single CSP. However, if $\langle\psi\lambda\rangle=0$,
which is to say when the two spinors are $\psi$ and $\lambda$
are \emph{collinear}, the ratio $\tfrac{\langle\psi\xi\rangle}{\langle\lambda\xi\rangle}$ is simply their proportionality coefficient, which is Lorentz invariant (for the standard action
generated by $L_{\alpha\beta}$ and $\tilde L_{\dot\alpha\dot\beta}$).
We will come back to this point shortly, when discussing
$3$-pt amplitudes for which collinearity particularly important.

Assuming generic, non-collinear kinematics, one can add up such particular solutions determined 
by different spinors $\psi^{(l)}$, and as the same reasoning
applies to the tilde spinors, one ends up with 
\begin{equation}
\label{eq:amp_ratios}
    \mathcal{R}_\xi = \frac{i}{2}\sum_{k \in I_{\scriptscriptstyle\sf CSP}}
    \mu_k\,(r_k+\tilde r_k)\,,
    \qquad\text{with}\qquad\left\{
    \begin{aligned}
        r_k & := \frac{1}{\langle\lambda^{(k)}\xi\rangle}\,
        \sum_l a_{k,l} \frac{\langle\psi^{(l)}\xi\rangle}{\langle\psi^{(l)}\lambda^{(k)}\rangle}\,, \\
        \tilde r_k & := \frac{1}{[\tilde\lambda^{(k)}\tilde\xi]}\,
        \sum_m \tilde a_{k,m} \frac{[\tilde\psi^{(m)}\tilde\xi]}{[\tilde\psi^{(m)}\tilde\lambda^{(k)}]}\,,
    \end{aligned}
    \right.
\end{equation}
for any choice of coefficients $a_{k,l}$ and $\tilde a_{k,l}$
such that
\begin{equation}
    \sum_l a_{k,l} = 1 = \sum_m \tilde a_{k,m}\,,
    \qquad\forall k \in I_{\scriptscriptstyle\sf CSP}\,,
\end{equation}
with the sums running over combinations of spinors
$\psi^{(l)}$ and $\tilde\psi^{(m)}$ which are \emph{not collinear}
to $\lambda^{(k)}$ and $\tilde\lambda^{(k)}$ respectively,
and hence the $n$-pt amplitude takes the form
\begin{equation}\label{eq:amplitude_CSP}
    \mathcal{A}_\xi 
    = e^{-\mathcal{R}_\xi}\,\times\,\mathscr{A}\,,
\end{equation}
where $\mathscr{A}$ is an amplitude of usual (massive and massless) particles in the spectrum 
$I_{\scriptscriptstyle\sf DSP}$, together with massless particles
of arbitrary helicities for the spectrum $I_{\scriptscriptstyle\sf CSP}$. Note that two different sets of coefficients $a_{k,l}$ and $\tilde a_{k,l}$ define two different particular solutions of the equations \eqref{eq:exponent}, and therefore differ by a solution of the homogeneous equation. In this sense,  the choice of these coefficients is tied to the possible ambiguity left in the solution of $L_{\alpha\beta}\mathscr{A}=0=\tilde L_{\alpha\beta}\mathscr{A}$. In the case of 3-point amplitudes (see below) these parameters disappear in the prefactors, a feature tied to the momentum conservation of three particles.
A similar factorization of the on-shell scattering amplitudes with continuous-spin legs was already observed in \cite[Section 3.3]{Bellazzini:2024dco} where a pair of complex spinors was used instead. In the light-cone gauge a similar structure was found in \cite{Metsaev:2017cuz,Metsaev:2018moa}.

%==============================%
\paragraph{Infinite-spin limit.}
%==============================%
As mentioned in the introduction, a continuous-spin particle, 
as well as the associated wave equation, can be obtained
from massive particle of mass $m$ and spin $J$ by taking
the simultaneous limit $m\to0$ and $J\to\infty$, with the product
$mJ \to \mu$ fixed \cite{Khan:2004nj, Bekaert:2005in}.
In the appendix, we argue that taking this limit
at the level of amplitudes allows one to recover the scattering amplitudes of processes involving one CSP and a number of massive/massless particles of discrete spin, from those amplitudes where the CSP leg is replaced by a massive one.

Start with an amplitude where one leg is a massive spin-$J$
particle, with momentum spinors $\lambda^I$, and all other legs
are either massive or massless of helicity-type (i.e. no CSP
is present for the moment). Such an amplitude takes the form
of a momentum-conserving Dirac distribution, times a function
depending on angle or square bracket of the momentum spinors.
The dependency of the latter function on the momentum spinors
$\lambda^I$ of the massive spin-$J$ particle is such that
\begin{equation}
\mathscr{A}\big(t\langle\lambda^I\psi^{(i)}\rangle,
    t[\tilde\lambda_I{\tilde\psi^{(j)}}],\dots\big)
    = t^{2J}\,\mathscr{A}\big(\langle\lambda^I{\psi^{(i)}}\rangle,
    [\tilde\lambda_I{\tilde\psi^{(j)}}],\dots\big)\,,
\end{equation}
where $\psi^{(i)}$ and $\tilde\psi^{(j)}$ are any combinations of momentum spinors which are \emph{not collinear} to $\lambda_I^{(k)}$ and $\tilde\lambda_I^{(k)}$ respectively,
and the dots denote other contractions of spinors not involving
$(\lambda^I, \tilde\lambda_I)$. In principle, the amplitude
carries $SU(2)$ indices, and in our situation, will have
at least a group of $2J$ symmetric indices corresponding
to those carried out by $\lambda^I$ and $\tilde\lambda_I$.
We can contract them by introducing an $SU(2)$ doublet $u_I$,
so that the dependency on the massive spin-$J$ momentum spinors
of the above amplitude appears through the combinations
$u_I \langle \lambda^I \psi^{(i)}\rangle$ 
and $u^I[\tilde\lambda_I \tilde\psi^{(j)}]$.
Assuming that $\mathscr{A}$
is a rational function of the spinor brackets, it can be written
as a linear combination of terms with \emph{definite homogeneity}
in each contraction involving $\lambda^I$ and $\tilde\lambda_I$,
i.e. 
\begin{equation}
    \mathscr{A} = \sum_{\sum_i b_i = 1 = \sum_j \tilde b_j}
    \mathscr{A}_{\{b_i,\tilde b_j\}}\,,
\end{equation}
where $\mathscr{A}_{\{b_i,\tilde b_j\}}$ is homogeneous
of degree $b_i J$ in 
$u_I\,\langle\lambda^I\psi^{(i)}\rangle$
and of degree $\tilde b_j J$ 
in $u^I\,[\tilde\lambda_I\tilde\psi^{(j)}]$.
In other words, $b_i$ are the fractions that measure the amount of the massive spinors $\lambda^I$ contracted onto the number $i$ particle. These fractions are to be held fixed in the limit $J\to\infty$. As expected, taking the infinite-spin limit requires a family of amplitudes. Introducing an auxiliary spinor $\xi$ such that
$\langle\lambda\xi\rangle\neq 0$ as before, one should set
$\lambda^{I=1}_\alpha=\lambda_\alpha$
and $\lambda^{I=2}_\alpha 
    =\tfrac{m}{\langle\lambda\xi\rangle}\xi_\alpha$,
to prepare for taking the massless limit. 
Furthermore, let us set $u_1=1$, and $u_2=i$, so that 
the spinor contraction take the form
\begin{equation}
    u_I\langle\lambda^I{\psi}\rangle\ 
    \leadsto\ \langle\lambda{\psi}\rangle\times
    \Big(1+i\,\tfrac{m}{\langle\lambda\xi\rangle}
    \tfrac{\langle\xi{\psi}\rangle}{\langle\lambda{\psi}\rangle}\Big)
    \qquad\text{and}\qquad
    u^I[\tilde\lambda_I{\tilde\psi}]\
    \leadsto\ i\,[\tilde\lambda{\tilde\psi}] \times
    \Big(1+i\,\tfrac{m}{[\tilde\lambda\tilde\xi]}
    \tfrac{[\tilde\xi{\tilde\psi}]}{[\tilde\lambda{\tilde\psi}]}\Big)\,.
\end{equation}
We can now take the double-scaling limit $m\to0$ and $J\to\infty$
with $mJ = \mu$ fixed on the homogeneous components, which yields
\begin{align}
    & \mathscr{A}_{\{b_i,\tilde b_j\}}
    \big(u_I\langle\lambda^I{\psi^{(i)}}\rangle,
    u^I[\tilde\lambda_I{\tilde\psi^{(j)}}],\dots\big)
    = \prod_{i,j} \Big(1+\tfrac{i}{J}
    \tfrac{mJ}{\langle\lambda\xi\rangle}
    \tfrac{\langle\xi{\psi^{(i)}}\rangle}{\langle\lambda{\psi^{(i)}}\rangle}\Big)^{J\,b_i}
    \Big(1+\tfrac{i}{J}\tfrac{mJ}{[\tilde\lambda\tilde\xi]}
    \tfrac{[\tilde\xi{\tilde\psi^{(j)}}]}{[\tilde\lambda{\tilde\psi^{(j)}}]}\Big)^{J\,\tilde b_j} \\ & \hspace{300pt} \nonumber 
    \times \mathscr{A}_{\{b_i,\tilde b_j\}}
    \big(\langle\lambda{\psi^{(i)}}\rangle,
    [\tilde\lambda{\tilde\psi^{(j)}}],\dots\big) \\
    & \hspace{50pt} 
    \underset{\substack{m\to0\\J\to\infty}}{\longrightarrow}
    \exp\Big(i\mu\,\sum_{i,j}
    \tfrac{b_i}{\langle\lambda\xi\rangle}
    \tfrac{\langle\xi{\psi^{(i)}}\rangle}{\langle\lambda{\psi^{(i)}}\rangle}
    +\tfrac{\tilde b_j}{[\tilde\lambda\tilde\xi]}
    \tfrac{[\tilde\xi{\tilde\psi^{(j)}}]}{[\tilde\lambda{\tilde\psi^{(j)}}]}\Big)
    \times \mathscr{A}\big(\langle\lambda{\psi^{(i)}}\rangle,
    [\tilde\lambda{\tilde\psi^{(j)}}],\dots\big)\,,
\end{align}
which reproduces the form of the amplitudes \eqref{eq:amp_ratios}-\eqref{eq:amplitude_CSP} in the case of only one CSP leg (i.e. $|I_{\scriptscriptstyle\sf CSP}|=1$). Note that $\mathscr{A}(\langle\lambda{\psi^{(i)}}\rangle,[\tilde\lambda{\tilde\psi^{(j)}}],\dots)$
should be understood
as an amplitude between the original spectrum
of particles, with the massive spin-$J$ one being replaced
by a massless particle of `unfixed helicity' (i.e. one sums
over all possible values for this helicity). Indeed, the infinite-spin limit with fractions $b_i$ and $\tilde b_j$ fixed means that the family of amplitudes runs over infinitely many values of spins of other particle. This latter
replacement is justified at the level of single representations
in the appendix, 
by studying the infinite-spin limit
of the spinor-helicity representation of a massive particle
to a continuous-spin one. Lastly, let us comment that this exponentiation in the infinite-spin limit is very similar to how the infinite-spin limit reproduces the scattering of black holes \cite{Guevara:2018wpp}.

%******************************************%
\subsection{Explicit examples of amplitudes}
\label{Explicitexamples}
%******************************************%
In this section we provide some explicit examples of amplitudes involving continuous-spin particles obtained from the general formula~\eqref{eq:amplitude_CSP}. Starting with three-point amplitudes, we consider ``functional'' solutions (namely, analytical functions of the kinematical variables) and find exact agreement with the amplitudes constructed in~\cite{Bellazzini:2024dco}. Yet, inspired by the distributional off-shell vertices found in \cite{Metsaev:2025wcv}, we consider three-point amplitudes with distributional support. 
With this extension we find distributional solutions in precisely the same cases as~\cite{Metsaev:2025wcv} which allow to circumvent some of the no-go's reported in~\cite{Bellazzini:2024dco}. In principle, all solutions directly emerge by imposing the Lorentz invariance. 

As far as distributional solutions are concerned, firstly, we find non-trivial distributional amplitudes between three continuous-spin particles, as well as between two continuous-spin particles and a massless helicity particle. In particular, we find that a coupling between a graviton and two continuous-spins is allowed in this distributional sense (which is an important feature if one wants to have a chance evading a CSP analogue of the no-go theorem \cite{Porrati:2008rm}, see also \cite{Bekaert:2010hw}). On the contrary, no distributional solutions seem to exist for the case of one continuous-spin and two massless helicity particles (which prevents a CSP to be a candidate IR-deformed graviton, in the sense that a CSP cannot reduce to an on-shell graviton in the UV limit \cite{Bellazzini:2024dco}). 
Secondly, the same arguments can be used to define a distributional solution for the amplitude between a CSP and two massive particles in the equal mass limit, which does not exist when restricting to functional solutions.

Moreover, we consider the possibility of chiral solutions to the Poincaré invariance constraints by trivializing the action of the Lorentz generators in one chirality sector. This allows to obtain functional solutions for massless three-point amplitudes including continuous-spin particles, at the cost of them being inherently complex.

At higher points we discuss the procedure to glue three-point amplitudes involving continuous-spin particles in order to construct four-point amplitudes, and provide explicit map between any functional $n$-point amplitude computed using the representation of continuous-spin particles introduced in this work and the ones obtained in~\cite{Bellazzini:2024dco}. We end this section by proposing some distributional $n$-point amplitudes based on the lessons from lower-point examples.

%************************************%
\subsubsection{Three-point amplitudes}
%************************************%
\paragraph{Functional solutions:} We start by discussing functional solutions to eq.~\eqref{eq:amplitude_CSP}, namely those given by analytical functions of the kinematical variables.

\vspace{1mm}

{\it One continuous-spin particle and two massive particles:} 
Consider the amplitude between two massive particles with spin $J_i$, mass $m_i$ and momentum $p_i$, with $i=1,2$ and a continuous-spin particle of parameter $\mu$, momentum $p_3$ and reference spinor $\xi_\alpha$.

The first step to apply formula~\eqref{eq:amplitude_CSP} is to construct all the independent ratios of spinor brackets that contribute to the exponential. Specializing eq.~\eqref{eq:amp_ratios} to the present case, we have
\begin{equation}
\label{eq:rations3ptnoneqmass}
    r=\frac{1}{\langle 3 \xi \rangle}\left(a \frac{\langle\psi^{(1)}\xi\rangle}{\langle\psi^{(1)}3\rangle}+(1-a)\frac{\langle\psi^{(2)}\xi\rangle}{\langle\psi^{(2)}3\rangle}  \right),
\end{equation}
where $a$ is an arbitrary constant. There is a similar expression for $\tilde{r}$. 

Yet, using the definition \eqref{definitionpsi} of $\psi^{(i)}_\alpha$ for the massive case, we have $\psi^{(i)}_\alpha = |i\rangle_{\alpha}^I\,[ i_{I\,}3]=p_{i\,\alpha \dot{\alpha}}|3]^{\dot{\alpha}}$, and thus
\begin{equation}
    \frac{\langle\psi^{(1)}\xi\rangle}{\langle\psi^{(1)}3\rangle}=\frac{[3|p_1|\xi\rangle}{[3|p_1|3\rangle}=\frac{[3|p_2|\xi\rangle}{[3|p_2|3\rangle}=\frac{\langle\psi^{(2)}\xi\rangle}{\langle\psi^{(2)}3\rangle},
\end{equation}
where we used momentum conservation and the fact that $p_{3\alpha \dot{\alpha}}|3]^{\dot{\alpha}}=0$. We see that there is only one independent ratio for this amplitude, and the dependence on the arbitrary constant $a$ disappears from the exponent. 

The general answer for the amplitude corresponding to this process is therefore given by
\begin{equation}
\label{eq:1CSP2massive}
   \mathcal{A}_\xi(1^{J_1},2^{J_2},3^\mu)=e^{-\tfrac{i\mu}{2}\left(\tfrac{1}{\langle 3 \xi \rangle} \frac{[3|p_1|\xi\rangle}{[3|p_1|3\rangle}+\frac{1}{[3 \tilde\xi ]}\frac{[\xi|p_1|3\rangle}{[3|p_1|3\rangle}\right)} \mathscr{A}(1^{J_1},2^{J_2},3)\,,
\end{equation}
where $\mathscr{A}(1^{J_1},2^{J_2},3)$ is a function of the momentum spinors, without any definite homogeneity in $|3\rangle_\alpha$. As shown in the previous section, it can be expressed as a sum of definite homogeneity contributions corresponding to the expansion of the continuous-spin state in the helicity basis, yielding
\begin{equation}
\label{eq:3ptCSP1m2}
   \mathcal{A}_\xi(1^{J_1},2^{J_2},3^\mu)=e^{-\tfrac{i\mu}{2}\left(\tfrac{1}{\langle 3 \xi \rangle} \frac{[3|p_1|\xi\rangle}{[3|p_1|3\rangle}+\frac{1}{[3 \tilde\xi ]}\frac{[\xi|p_1|3\rangle}{[3|p_1|3\rangle}\right)} \sum_{h \in \mathbb{Z}}\mathscr{A}(1^{J_1},2^{J_2},3^h)
   \,,
\end{equation}
where each term in the sum corresponds to a scattering amplitude between the two massive particles and a massless particle of helicity $h$, which were fully classified in~\cite{Conde:2016vxs, Conde:2016izb, Arkani-Hamed:2017jhn}. 

In~\cite{Bellazzini:2024dco}, the authors argued that three-point amplitudes between a continuous-spin particle and two particles of degenerate mass do not exist (see also \cite{Bekaert:2017xin} for an earlier observation of this property in the case of two massive scalars), dubbing this property ``mass-splitting selection rule'', as the equal mass limit of the amplitude with generic masses is badly singular. We can confirm this feature in our formulation by noting that $[3|p_1|3\rangle=2 p_1 \cdot p_3=m_2^2-m_1^2$, and therefore the equal mass limit of~\eqref{eq:3ptCSP1m2} is ill-defined, as the amplitude exhibits an essential singularity of the form $e^{\mu/(m_2^2-m_1^2)}$. Nevertheless, in agreement with the observation by Metsaev (see \cite{Metsaev:2025wcv} and refs therein) one should stress that that the amplitude does not vanish in the degenerate mass case if we allow for \textit{distributional} amplitudes (cf. the comments at the end of this subsection).

\vspace{1mm}

{\it Two continuous-spin particles and one massive particle:} Consider now the case with two continuous-spin particles of momentum $p_1$ and $p_2$, and a massive particle of spin $J$ and momentum $p_3$. In this case, the factors $r_i$ in the exponential are
\begin{equation}
    \begin{split}
        r_1=\frac{1}{\langle 1 \xi \rangle}\left( a_1 \frac{\langle2\xi\rangle}{\langle21\rangle}+(1-a_1) \frac{[1|p_3|\xi\rangle}{[1|p_3|1\rangle}  \right),\\
         r_2=\frac{1}{\langle 2 \xi \rangle}\left( a_2 \frac{\langle1\xi\rangle}{\langle1 2\rangle}+(1-a_2) \frac{[2|p_3|\xi\rangle}{[2|p_3|2\rangle}  \right),  
    \end{split}
\end{equation}
with $a_1$ and $a_2$ arbitrary constants. Yet, once again the ratios are not independent, as $[1|p_3|1\rangle=[2|p_3|2\rangle=-m^2$, $m^2=\langle12\rangle[12]$ and momentum conservation imply
\begin{equation}
   \frac{[1|p_3|\xi\rangle}{[1|p_3|1\rangle} =\frac{[1|p_2|\xi\rangle}{m^2}=\frac{[12]\langle2\xi\rangle}{m^2}=\frac{\langle2\xi\rangle}{\langle21\rangle},
\end{equation}
and similarly for $r_2$. As in the previous case, the dependence on the arbitrary constants drops, and we obtain that the amplitude for this process is then given by
\begin{equation}
\label{eq:3ptCSP2m1}
   \mathcal{A}_{\xi}(1^{\mu_1},2^{\mu_2},3^J)=e^{-\frac{i\mu_1}{2}\left(\frac{\langle2\xi\rangle}{\langle 1 \xi \rangle\langle21\rangle}+\frac{[2\xi]}{[1\xi][21]}\right)} e^{-\frac{i\mu_2}{2}\left(\frac{\langle1\xi\rangle}{\langle 2 \xi \rangle\langle 12 \rangle}+\frac{[1\xi]}{[ 2 \xi][12]}\right)} \sum_{h_1,\, h_2 \, \in \mathbb{Z}} \mathscr{A}(1^{h_1},2^{h_2},3^J),
\end{equation}
where the sum is over amplitudes between two massless particles of helicities $h_1$ and $h_2$ and a massive spin $J$ particle.

We can check that the massless limit of the amplitude is singular by noting that $m^2=\langle12\rangle[12]$ causes either the terms proportional to $\langle12\rangle^{-1}$ or the ones proportional to $[12]^{-1}$ to diverge when $m \to 0$, producing an essential singularity of the same kind as in the previous example. Yet again, allowing for distributionally-supported amplitudes permits to find non-trivial solutions also in this case, as will be discussed below. Another possible way out is to consider chiral vertices where either $[12]$ or $\langle12\rangle$ is absent, see the discussion on chiral amplitudes below.

\vspace{1mm}

{\it Three massless particles, with at least one of continuous-spin type:} For a three-point function involving only massless particles
with at least one continuous-spin one, we run into a kinematical
obstruction due to Lorentz invariance. Indeed, as is well-known for three massless particles 
one cannot have momentum conservation \emph{and} non-collinear momenta
unless one complexify them and require all square-brackets
(or all angle-brackets) to vanish identically (since it follows from $p_i\cdot p_j=0$ that $[ij]\langle ij\rangle=0$). For massless particles
of helicity-type, this is compatible with Lorentz invariance, 
but for the continuous-spin case, the general expression
\eqref{eq:amplitude_CSP} shows that we need both types
of spinor contractions---angle and square brackets---in order
to ensure that the amplitude be Lorentz-invariant.

Typically, we need to make sense of 
\begin{equation}
    \delta\big([12]\big)\,
    \delta\big([23]\big)\,
    \exp\Big(\!-\tfrac{i\mu}{2}\tfrac{[2\tilde\xi]}{[1\tilde\xi][12]}\Big)
    \times \big(\dots\big)\,,
\end{equation}
as a distribution, where we assumed that the spinor $|1]$
is that of the CSP with parameter $\mu$
and reference spinor $|\tilde\xi]$, and $|2]$ is another
massless particle. The overall product of Dirac delta's
comes from momentum conservation, and impose that 
\begin{equation}
    |1] \propto |2] \propto |3]\,,
\end{equation}
so that in particular $|2\tilde\xi] = \alpha\,|1\tilde\xi]$
for some constant $\alpha \in \mathbb{C}$, and therefore 
the above expression boils down to
\begin{equation}
    \delta([12])\,\exp\Big(-\tfrac{i\mu}{2}\tfrac{\alpha}{[12]}\Big)
    \times \big(\dots\big)\,
\end{equation}
which exhibits an essential singularity and thus cannot represent a well-defined scattering amplitude.
This agrees with the findings of \cite{Metsaev:2025wcv} and \cite{Bellazzini:2024dco},
wherein the authors observed that, respectively, cubic vertices and $3$-pt amplitudes
between massless fields with at least one CSP are not allowed. More precisely, this no-go result relies on two assumptions which are made explicit in the classification \cite{Metsaev:2025wcv}: (1) parity symmetry is assumed and (2) one does not allow for distributional functions of the kinematical space. If one of these two assumptions is relaxed, then solutions can be found.

\paragraph{Chiral solutions:} Notice that the above obstruction arises if one insists
on imposing invariance under $SL(2,\mathbb{C})$ together with the reality condition that relates the $SL(2,\mathbb{C})$-actions on left and right spinors. This problem arises first for a scattering of three massless helicity particles and is still present if any of them are replaced by continuous-spin particles. Indeed, let us add the momentum conservation to the most general (holomorphic) three-point amplitude of massless helicity fields. Remember that $p_i\cdot p_j=0$ is equivalent to $[ij]\langle ij\rangle=0$. Therefore, in the real Minkowski kinematics any three-point amplitude---except for $\mathscr{A}_{0,0,0}$---must vanish. As it was mentioned, a way to have nontrivial solutions to $[ij]\langle ij\rangle=0$ is to complexify the momentum space and, hence, to have two independent copies
of $SL(2,\mathbb{C})$ making up the complexification
of the (double cover of the) Lorentz group,
i.e. those generated by $\lambda_\alpha$
and by $\tilde\lambda_{\dot\alpha}$. Alternatively, one can pass to the split signature where $|i]$ and $|i\rangle$ are independent. 

Now, imposing the amplitude to be invariant under translation---that is requiring momentum conservation---as well as
under the left copy of $SL(2,\mathbb{C})$ whose action
is generated by $J_{\alpha\beta}$ and the right copy whose action is generated by $\tilde J_{\dot\alpha\dot\beta}$, does allow for a solution:
the amplitude should simply be holomorphic, and therefore
takes the form
\begin{equation}
    \mathcal{A}_\xi(1^\mu,2^{h_2},3^{h_3})
    = e^{\frac{i\mu}{2}\frac{\langle2\xi\rangle}{\langle1\xi\rangle\langle12\rangle}}\,
    \mathscr{A}(1^\mu,2^{h_2},3^{h_3})\,,
\end{equation}
with $\mathscr{A}$ a $3$-pt amplitude for massless fields
with only two of them having fixed helicity $h_2$ and $h_3$.
Such a solution is consistent with the fact that ordinary,
helicity-type, massless particles are also representations
of a `chiral' or `deformed' version of the Poincar\'e algebra,
as noticed in \cite{Ponomarev:2022ryp, Ponomarev:2022qkx}.
The latter is defined as the algebra generated by translations
$P_{\alpha\dot\alpha}$, one copy of $SL(2,\mathbb{C})$%
---say $J_{\alpha\beta}$---and central generators 
$C_{\dot\alpha\dot\beta}$ such that $[P_{\alpha\dot\alpha},P_{\beta\dot\beta}]
\propto \epsilon_{\alpha\beta}C_{\dot\alpha\dot\beta}$,
i.e. deforming the commutation relation between translations
in the sector of opposite chirality with respect to
the $SL(2,\mathbb{C})$ copy kept. These central generators
can simply be represented trivially on the usual representation
for massless particles of helicity-type, and the same is true
of the continuous-spin representation we are focusing on here.

In other words, the usual $3$-pt massless amplitudes can be understood
as amplitudes for `chiral Poincar\'e particles', meaning
representation of the chiral version of the Poincar\'e group
discussed above, and from this viewpoint, continuous-spin particles
are on the same footing. Proceeding with this in mind, 
we can expand the previous expression with respect to helicities
of the continuous-spin particles, and find
\begin{align}
\label{eq:amplitude_CSP_helicities}
    \mathcal{A}_\xi(1^\mu,2^{h_2},3^{h_3})
    & = e^{\frac{i\mu}{2}
    \frac{\langle2\xi\rangle}{\langle1\xi\rangle\langle12\rangle}}\,\langle12\rangle^{2h_2}\langle13\rangle^{2h_3}
    \sum_{h \in \mathbb{Z}} c_h\,\bigg[\frac{\langle23\rangle}{\langle12\rangle\langle13\rangle}\bigg]^h \\
    & = e^{\frac{i\mu}{2}
    \frac{\langle2\xi\rangle}{\langle1\xi\rangle\langle12\rangle}}\,
    \sum_{h \in \mathbb{Z}} \tilde c_h\,
    \mathscr{A}_{h,h_2,h_3}(1^h,2^{h_2},3^{h_3})\,,
\end{align}
where the coefficients $c_h$ and $\tilde c_h$
are a priori arbitrary, and
\begin{equation}
    \mathscr{A}_{h,h_2,h_3}(1^{h},2^{h_2},3^{h_3})
    = \langle12\rangle^{h+h_2-h_3}\langle13\rangle^{h-h_2+h_3}
    \langle23\rangle^{-h+h_2+h_3}\,,
\end{equation}
are the usual `elementary' $3$-pt massless amplitudes, 
between particles of helicities $h$, $h_2$ and $h_3$.

%=====================================================%
\paragraph{(Super-)collinearity of massless particle amplitudes:}
%=====================================================%
In the same spirit of holomorphic solutions, let us illustrate an additional effect of having an infinite multiplet of massless helicity fields. Let us assume that $[ij]\neq0$. Now, the momentum conservation can be rewritten to manifest $\langle ij\rangle=0$
\begin{equation}\label{collinearity}
   \delta^{(4)}\left(\sum_{i=1}^3\lambda^{(i)}_{\alpha}\tilde \lambda^{(i)}_{\dot \alpha}\right)\,=\, \frac{\langle q3 \rangle^2 }{\big|\,[12]\,\big|}\, \delta\big(\langle13 \rangle\big) \,\delta\big( \langle32\rangle\big)\, \delta^{(2)}  \Big(\sum_{i=1}^3 \langle qi\rangle [i|_{\dot\alpha} \Big)\,,
\end{equation}
where $|q\rangle_\alpha$ is a reference spinor such that $\langle q3 \rangle \neq 0$.
As was shown in \cite{Ponomarev:2022atv}, in a theory with massless fields of all spins one can construct a generating function by summing over all three helicities $h_{1,2,3}$. Let us take the three-point amplitude of chiral higher-spin gravity for concreteness 
\begin{equation}\label{3ptmassless}
    \mathscr{A}_{h_1,h_2,h_3}(1^{h_1},2^{h_2},3^{h_3})
    = \frac{\ell_p^{h_1+h_2+h_3-1}}{\Gamma(h_1+h_2+h_3)}[12]^{h_1+h_2-h_3}[31]^{h_1-h_2+h_3}
    [23]^{-h_1+h_2+h_3}\,,
\end{equation}
where $\ell_p$ is a parameter with the dimension of length.
Note that one has to sum at the condition $h_1+h_2+h_3>0$. In order to perform the sum we use the standard regularization 
\begin{align}
    \sum_h z^h =\delta(1-z)
\end{align}
The final result is a ``super-collinear'' amplitude
\begin{align}\label{superduper}
e^{[12]\ell_p}\,\delta^{(2)}(\bar{1}^{\dot{\alpha}}+\bar{2}^{\dot{\alpha}}+\bar{3}^{\dot{\alpha}}) \delta^{(2)}( 1^\alpha-2^\alpha)\delta^{(2)}( 2^\alpha-3^\alpha)\,.
\end{align}
Note that while for three massless particles of fixed helicities the spinors $|i\rangle$ are just parallel/collinear, i.e. $\langle ij\rangle=0$, after the summation over all helicities the generating function has factors $\delta^{(2)}(|i\rangle- |j\rangle)$, i.e. the spinors are not just collinear but are identical: they are ``super-collinear''. This does not mean that the scattering of individual massless helicity-type particles requires $|i\rangle= |j\rangle$, it is just a peculiar feature of having an infinite multiplet and a generating function of all amplitudes.\footnote{In other words, while collinearity of real momenta is a kinematical consequence of Poincar\'e symmetry, super-collinearity of the momentum spinors arises as a consequence of the higher-spin extension of Poincar\'e symmetries.} These observations will be useful below since amplitudes of continuous-spin particles are closer to those of an infinite multiplet of massless helicity-type particles.

%===================================%
\paragraph{Distributional solutions:}
%===================================%
When considering functional solutions to Poincaré invariance in the previous section, we found that the general solution~\eqref{eq:amplitude_CSP} breaks down for three-point amplitudes between massless particles involving at least one continuous-spin particle, as well as a single continuous-spin particle and two massive particles of equal mass, as reported in~\cite{Bellazzini:2024dco}. 

We can trace the origin of this failure to the fact that the ratios~\eqref{eq:amp_ratios} that provide particular solutions to~\eqref{eq:exponent} for generic kinematics stop fulfilling this purpose in these degenerate cases. For example, as mentioned before, three-point massless kinematics requires all spinors of a given chirality to be collinear. Choosing these to be the holomorphic momentum spinors $|k\rangle_\alpha$ (with $k=1,2,3$), we see that the ratios $\langle i\xi \rangle\big/\langle j\xi \rangle$, which provide solutions to~\eqref{eq:exponent} in generic kinematics, rather satisfy 
\begin{equation}
    L_{\alpha \beta} \frac{\langle i\xi \rangle}{\langle j\xi \rangle}=\frac{\langle ij \rangle}{\langle j  \xi \rangle^2}\xi_\alpha \xi_\beta =0\,,
\end{equation}
in collinear kinematics, since $\langle ij \rangle=0$. This is to be expected, as in the collinear limit the above ratio yields simply the proportionality coefficient between the momentum spinors, called here ``collinear fraction'', which is Lorentz invariant.

Similar is the situation for the case of a CSP and two equal-mass massive particles: While for $m_1 \neq m_2$ we can define two independent spinors made out of the momentum variables to span $\mathbb{C}^2$ given by $\lambda^{(3)}_{\alpha }$ and $\tfrac{1}{m_1}p_{1 \alpha \dot\alpha} \tilde \lambda^{(3)\dot\alpha}$ (with $p_1$ the momentum of one of the massive particles and $\{\lambda^{(3)},\tilde \lambda^{(3)}\}$ the momentum spinors associated to the CSP), these become collinear in the equal mass limit, causing the solutions obtained for generic masses to break down.

Yet, this does not necessarily mean that solutions do not exist in these special kinematical configurations, but rather that they must be treated separately. As we will show below, it is possible to find non-trivial three-point amplitudes in these cases, with the particularity that these are distributional: In order to be consistent, these amplitudes have support on kinematical configurations in which the spinors are not merely collinear, but their proportionality coefficients, the collinear fractions, are constrained by the continuous-spin scales of the CSP present in the process. This is in perfect agreement with the findings of~\cite{Metsaev:2017cuz,Metsaev:2018moa,Metsaev:2025wcv}, where the author reported the existence of distributional solutions for off-shell cubic vertices in the light-cone gauge involving continuous-spin particles in cases in which functional solutions do not exist. 

\vspace{1mm}

{\it One continuous-spin particle and two massive particles of equal mass:} As explained previously,~\eqref{eq:1CSP2massive} does not provide a solution in the equal mass limit of the three-point amplitude between a CSP and two massive particles of arbitrary mass is singular, as $\langle 3 | p_1|3]$ vanishes.
Yet, for these specific kinematics we can construct other combinations of spinor brackets that allow to find non-trivial solutions to~\eqref{eq:exponent}. 

Consider first the holomorphic equation for the exponent. Solving~\eqref{eq:exponent} requires finding a combination $r$ of spinor brackets that satisfies
\begin{equation}
\label{eq:Leqmass}
    L_{\alpha \beta } \, r=\frac{i\mu}{2}\xi_\alpha \xi_\beta\,.
\end{equation}
Apart from the combinations used in~\eqref{eq:rations3ptnoneqmass}, in equal mass kinematics we can consider the angle brackets $\langle 3 \xi \rangle$, $\langle \chi \xi \rangle$, where $\chi_\alpha = p_{1 \alpha \dot \alpha} \tilde \xi^{\dot \alpha}$, which transform under $L_{\alpha \beta}$ as 
\begin{equation}
    L_{\alpha \beta } \, \langle 3 \xi \rangle=- \lambda^{(3)}_{(\alpha} \xi_{\beta)}, \quad  L_{\alpha \beta } \, \langle \chi \xi \rangle=- \chi_{(\alpha} \xi_{\beta)}\,.
\end{equation}
Since $\chi_\alpha$ and $\lambda^{(3)}_\alpha$ are not collinear, we can expand $\xi_\alpha$ in the basis of $\mathbb{C}^2$ they define and find a combination $r$ of  $\langle 3 \xi \rangle$ and $\langle \chi \xi \rangle$ that satisfies~\eqref{eq:Leqmass}:
\begin{equation}
   L_{\alpha \beta } \, r=\frac{i\mu}{2}\xi_\alpha \xi_\beta, \quad {\rm for} \, \, \,r:= \frac{i\mu}{2} \frac{\langle \chi \xi\rangle}{\langle 3 \xi \rangle \langle 3 \xi \rangle}=\frac{i\mu}{2} \frac{\langle \xi | p_1|\tilde \xi]}{\langle 3|p_1|\tilde \xi ] \langle 3 \xi \rangle}\,,
\end{equation}
where, in the second equality, we used the definition of $\chi_\alpha$.

However, this ratio transforms non-trivially under the antiholomorphic Lorentz generator $\tilde{L}_{\dot \alpha \dot \beta}$, and therefore for it to be a consistent exponent function $\mathcal{I}_\xi$ we must ensure that it satisfies as well 
\begin{equation}
    \tilde{L}_{\dot\alpha \dot\beta } \, r=\frac{i\mu}{2}\tilde{\xi}_{\dot\alpha} \tilde{\xi}_{\dot\beta}\,.
\end{equation}
Using the same reasoning, it is simple to show that the antiholomorphic equation is satisfied by $\tilde r$:
\begin{equation}
     \tilde{L}_{\dot\alpha \dot\beta } \, \tilde r=\frac{i\mu}{2}\tilde{\xi}_{\dot\alpha} \tilde{\xi}_{\dot\beta}\,, \qquad \qquad \tilde r= \frac{i\mu}{2} \frac{\langle \xi | p_1|\tilde \xi]}{\langle \xi|p_1|3 ] [ 3 \tilde\xi ]} \,,
\end{equation}
hence the holomorphic and antiholomorphic equations can only be satisfied simultaneously if $r=\tilde r$, which implies
\begin{equation}
    \langle \xi|p_1|3 ] [ 3 \tilde\xi ]=\langle 3|p_1|\tilde \xi ] \langle 3 \xi \rangle
    \qquad\Longleftrightarrow\qquad
    x = \tilde x\,,
\end{equation}
where $x$ is a factor defined in~\cite{Arkani-Hamed:2017jhn} that gives to the proportionality constant between the spinors $|3\rangle_\alpha$ and $\tfrac{1}{m}p_{1\alpha \dot \alpha }|3]^{\dot \alpha}$ when they become collinear in the equal-mass kinematical configuration:
\begin{equation}
    x |3\rangle_\alpha = \frac{p_{1 \alpha \dot\alpha}}{m} |3]^{\dot \alpha}
    \qquad\Longrightarrow\qquad
    x:=\frac{[3|p_1|\xi\rangle}{m\langle  \xi 3 \rangle}\,, \qquad {\rm for } \, \, m_1=m_2=m\,.
\end{equation}

Moreover, on-shell three-point kinematics enforces $\tilde x = x^{-1}$, and as a consequence the solution involving the ratio $r$ is supported for kinematics in which $x^2=1$, namely $|3 \rangle_\alpha = \pm \tfrac{1}{m}p_{1\alpha \dot \alpha }|3]^{\dot \alpha}$. We see thus that the presence of a CSP restricts the already constrained three-point kinematics even more: the spinors $|3\rangle_\alpha$ and $p_{1 \alpha \dot\alpha}|3]^{\dot \alpha}$ are constrained to be not merely collinear as required by on-shell three-point kinematics, but they must be either identical or opposite. This is very similar to the generating function of three-point amplitudes of chiral highers-spin gravity \eqref{superduper} reviewed above. We will find this phenomenon again in what follows, when we consider massless three-point amplitudes including continuous-spin particles.

Collecting the previous results we obtain a non-trivial, distributionally-supported three-point amplitude between a continuous-spin particle and two massive particles of equal mass:
\begin{equation}
    \label{eq:1CSP2massiveeqmass}
    \mathcal{A}_\xi=\delta(x^2-1) \exp \left(-\frac{i\mu}{2} \frac{\langle \xi | p_1|\tilde \xi]}{\langle 3|p_1|\tilde \xi ] \langle 3 \xi \rangle}  \right) \mathscr{A}\,,
\end{equation}
where as before $\mathscr{A}$ is simply a function of angle and square brackets between the momentum spinors carrying the appropriate $SU(2)$ representations corresponding to the spin of the massive particles. Note that since the spin of the massive particles played no role in the construction, the above formula applies to arbitrary spin. Lastly, let us add that if the actions on right and on left spinors are kept independent, i.e. we consider chiral solutions, the relevant Dirac distribution is $\delta(x \mu - \tilde{x} \tilde{\mu})$ instead, where $\tilde{\mu}$ is the continuous-spin scale of the antichiral sector (assuming that they are independent).

\vspace{1mm}

{\it Three continuous-spin particles:}  Without loss of generality, let us consider three-point massless kinematics in the antiholomorphic configuration, namely the one in which all angle brackets are vanishing and thus holomorphic momentum spinors are collinear.
The momentum-conserving delta-function contained in the amplitude can be written in a way that makes the collinearity of holomorphic spinors in this configuration manifest
\begin{equation}
   \delta^{(4)}\left(\sum_{i=1}^3\lambda^{(i)}_{\alpha}\tilde \lambda^{(i)}_{\dot \alpha}\right)\,=\,\frac{\langle q3 \rangle^2 }{\big|\,[12]\,\big|}\, \delta\big(\langle13 \rangle\big)\, \delta\big(\langle32\rangle\big)\, \delta^{(2)}  \Big(\sum_{i=1}^3 \langle qi\rangle [i|_{\dot\alpha} \Big)\,,
\end{equation}
where $|q\rangle_{\alpha}$ is a reference spinor such that $\langle q 3\rangle \neq 0$. 

In this kinematical configuration, the only non-trivial spinor contractions we can build are contractions of a momentum spinor $\lambda^{(k)}_\alpha$ and $\xi_\alpha$. As was explained above, their ratios are annihilated by $L_{\alpha \beta}$. Therefore, in this case it is not possible to find solutions to the inhomogeneous equation~\eqref{eq:exponent}. Instead, to obtain non-trivial solutions for three-point amplitudes between three continuous-spin particles we must consider solutions to the (holomorphic) Lorentz invariance constraints $J_{\alpha \beta} \mathcal{A}=0$ that are annihilated by $L_{\alpha \beta}$ and $U_{\alpha \beta}$, simultaneously, with $U_{\alpha \beta}$ the multiplicative part of the Lorentz generators:
\begin{equation}
    U_{\alpha \beta}=\sum_{k \in I_{\scriptscriptstyle\sf CSP}}
    \tfrac{\mu_k}{\langle\lambda^{(k)}\xi\rangle^2}\,
    \xi_\alpha\xi_\beta \,.
\end{equation}
Acting with $U_{\alpha \beta}$ on a candidate amplitude between three continuous-spin particles yields
\begin{equation}
    U_{\alpha \beta}\mathcal{A}_\xi=\left(  \frac{\mu_1}{\langle 1\xi\rangle^2}\,
    +\frac{\mu_2}{\langle 2\xi\rangle^2}\,
    +\frac{\mu_3}{\langle3\xi\rangle^2}
    \right) \xi_\alpha\xi_\beta \, \mathcal{A}_\xi\,,
\end{equation}
and thus the requirement $U_{\alpha \beta}\mathcal{A}_\xi=0$ implies that $\mathcal{A}_\xi$ is supported on configurations such that the prefactor involving the continuous-spin scales vanishes, namely
\begin{equation}
\label{eq:Uinvariance}
    \mathcal{A}_\xi=\delta\left(  \frac{\mu_1}{\langle 1\xi\rangle^2}\,
    +\frac{\mu_2}{\langle 2\xi\rangle^2}\,
    +\frac{\mu_3}{\langle3\xi\rangle^2}
    \right) \times \mathscr{A}\,,
\end{equation}
where $\mathscr{A}$ is an unconstrained function of the square brackets between the antiholomorphic momentum spinors.

Since in the antiholomorphic configuration all holomorphic spinors are collinear, we can express~\eqref{eq:Uinvariance} in a manifestly Lorentz-invariant way by introducing the unit spinor $\omega_\alpha$ defining their common direction, and writing
\begin{equation}
    |i\rangle_\alpha=t_i  \,\omega_\alpha, \quad i=1,2,3\,,
\end{equation}
where $t_i$ are the collinear fractions. 
In terms of the $t_i$, the Dirac delta in eq.~\eqref{eq:Uinvariance} can be expressed equivalently by 
\begin{equation}
\label{eq:CSPdelta}
    \delta\left( \frac{\mu_1}{t_1^2}\,
    +\frac{\mu_2}{t_2^2}\,
    +\frac{\mu_3}{t_3^2}
    \right)\,,
\end{equation}
up to an irrelevant Jacobian $\langle \omega \xi \rangle^2$, 
and therefore restricts the support of the amplitude to configurations in which the holomorphic spinors are not only collinear but have proportionality coefficients $t_i$ related by the continuous-spin scales $\mu_i$ through the constraint $\tfrac{\mu_1}{t_1^2}+\tfrac{\mu_2}{t_2^2}+\tfrac{\mu_3}{t_3^2}=0$.

Since~\eqref{eq:CSPdelta} is manifestly Lorentz invariant, invariance under $L_{\alpha \beta}$ simply requires that $\mathscr{A}$ in~\eqref{eq:Uinvariance} depends on the holomorphic spinors via angle brackets only, without ever invoking the problematic exponentials with singular behavior in the collinear limit that arose in the above discussion for generic momenta.

Finally, we must impose Lorentz invariance in the antiholomorphic sector, namely $\tilde J_{\dot \alpha \dot \beta} \mathcal{A}=0$. Yet, antiholomorphic spinors are not constrained to be collinear and thus in this sector we can apply the general procedure from the previous section to obtain the same exponential factor but now involving square spinors only, which are free of singularities.

Including the Dirac deltas arising from momentum conservation and $U_{\alpha \beta}$ invariance we obtain the following non-trivial expression for the 3 CSP amplitude:

\begin{equation}
\label{eq:A3CSP}
\begin{split}
    \mathcal{A}_\xi\,= &\,\,\frac{\langle q3 \rangle^2 }{\big|\,[12]\,\big|}\, \delta\big(\langle13 \rangle\big)\, \delta \big(\langle32\rangle\big)\, \delta^{(2)}  \Big(\sum_{i=1}^3 \langle qi\rangle [i|_{\dot\alpha} \Big) \, \delta\left( \frac{\mu_1}{t_1^2}\,
    +\frac{\mu_2}{t_2^2}\,
    +\frac{\mu_3}{t_3^2}
    \right)\times \\&  \quad  \exp\Big(-\tfrac{i}{2}\Big[\mu_1\tfrac{[2 \tilde\xi]}{[ 1 \tilde\xi ][21]}+\mu_2\tfrac{[1 \tilde \xi]}{[ 2\tilde \xi][ 12]} +\mu_3\tfrac{[2 \tilde\xi]}{[ 3 \tilde\xi] [ 23]}\Big]\Big) \times \mathscr{A}\left(t_k, [ij]\right)\,.
\end{split}
\end{equation}
Here, one could leave out any one of the three collinear fractions from $\mathcal{A}$ as it can be expressed in terms of the other two.

\vspace{1mm}

{\it Two continuous-spin particles and a massless helicity particle:} We can deal with this case using the same approach as above. It turns out that it is smoothly connected to the case of three continuous-spin particles by setting the continuous-spin scale of one of them to zero. Choosing for instance the particle 3 to be a massless helicity state (and thus setting $\mu_3=0$), we find that~\eqref{eq:A3CSP} reduces to
\begin{equation}
\label{eq:A2CSP}
\begin{split}
    \mathcal{A}_\xi\,= &\,\,\frac{\langle q3 \rangle^2 }{\big|\,[12]\,\big|}\, \delta\big(\langle13 \rangle\big)\, \delta\big(\langle32\rangle\big)\, \delta^{(2)}  \Big(\sum_{i=1}^3 \langle qi\rangle [i|_{\dot\alpha} \Big) \, \delta\left( \frac{\mu_1}{t_1^2}\,
    +\frac{\mu_2}{t_2^2}
    \right) \times \\&   \quad \exp\Big(-\tfrac{i}{2}\Big[\mu_1\tfrac{[2 \tilde\xi]}{[ 1 \tilde\xi ][21]}+\mu_2\tfrac{[1 \tilde \xi]}{[ 2\tilde \xi][ 12]} \Big]\Big) \times \mathscr{A}\left(t_{1,2}, [ij]\right)\,.
\end{split}
\end{equation}
and thus enforces the support of the amplitude to configurations in which the angle spinors of the continuous-spin particles are related by a fixed proportionality constant: 
\begin{equation}
    |1 \rangle_\alpha = \pm i\,\sqrt{\frac{\mu_1}{\mu_2} }\,|2\rangle_\alpha\,.
\end{equation}

Remarkably, this implies that---contrary to previous expectations---continuous-spin particles can in fact gravitate, in the sense that they admit a non-vanishing three-point coupling to gravitons, albeit one with only distributional support.

\vspace{1mm}

{\it One continuous-spin particle and two massless helicity particles:} The presented strategy does not allow us to find non-trivial amplitudes for a single CSP and two massless particles, as in this case the invariance under $U_{\alpha \beta}$ induces the delta function $\delta(\mu_1)$ and thus constrains $\mu_1$ to vanish, yielding an amplitude between discrete spin particles as the only solution. This coincides with the results of~\cite{Metsaev:2025wcv}, where no distributional solutions where found for the cubic vertex in this case and it is clear now that no distributional solutions are possible.

%*****************************************************%
\subsubsection{Four point amplitudes and factorization}
%*****************************************************%
The mechanism to construct functional four-point amplitudes is the same as in the previous three-point examples: For a given set of external states involving continuous-spin particles, the four-point amplitude will factorize into an exponential determined by the continuous-spin scales $\mu_i$, the independent ratios $r_i$ and a function of momentum spinors only, which can in turn be decomposed as an infinite sum over amplitudes in which the continuous-spin particles are replaced by massless helicity particles of all possible helicities. Thus, rather than deriving even more explicit examples let us comment on how to deal with the factorization of a four-point amplitude at a pole in which a CSP is exchanged, which is a genuine new feature of four-point amplitudes.

Consider a four-point amplitude with arbitrary external states. At a pole associated to the exchange of a given on-shell intermediate state, the residue of the amplitude factorizes into the product of two three-point functions between a pair of external states and the exchanged particle, summed over the quantum numbers of the intermediate states. 

More concretely, consider as an example an $s$-channel pole in which a state $X$ of mass $M^2$ is exchanged. Schematically, the statement is that 
\begin{equation}
   \mathcal{A}_4(1,2,3,4) \sim \frac{1}{s-M^2}\sum_{\alpha}\mathcal{A}_3^{(L)}\big(1,2,X(p,\alpha)\big)\mathcal{A}_3^{(R)}\big(X(-p,\alpha),3,4\big),
\end{equation}
as $s\to M^2$. Here $\alpha$ labels a basis of the one-particle Hilbert space of the exchanged particle, and we explicitly denoted that the momentum of the intermediate state is opposite in each amplitude, as we consider all incoming particles by convention.

To implement the sum over the Hilbert space of the intermediate state when it is a CSP expressed in our representation, we expand the three-point amplitudes $\mathcal{A}^{(L)}_3$ and $\mathcal{A}^{(R)}_3$ in a basis of helicity eigenstates.\footnote{Naturally this requires $M^2=0$.} Since helicity also reverses sign when we pass from an outgoing to an incoming state, tracing over the CSP Hilbert space yields
\begin{equation}
 \underset{s \to 0}{\mathrm{Res}}\,\mathcal{A}_4=   \sum_{h \in \mathbb{Z}}\mathcal{A}_{3,h}^{(L)}(\lambda,\tilde{\lambda},\dots)\mathcal{A}_{3,-h}^{(R)}(\lambda,-\tilde{\lambda},\dots),
\end{equation}
where $\lambda$ and $\tilde{\lambda}$ are the momentum spinors associated to the CSP intermediate state, and $\mathcal{A}_{3,h}^{(L/R)}$ denotes the component of $\mathcal{A}_{3}^{(L/R)}$ with helicity $h$, namely the piece homogeneous in $\lambda$ with degree $-2h$. Note that we use the convention $\lambda(-p)=\lambda(p)$, $\tilde{\lambda}(-p)=-\tilde{\lambda}(p)$, and we omitted the labels associated to the external particles.

Each helicity component can be projected out from the three-point amplitudes by
\begin{equation}
    \mathcal{A}_{3,h}(\lambda,\dots)=\frac{1}{2\pi}\int_{0}^{2\pi}{\rm d}\theta \,e^{-ih\theta}\mathcal{A}_{3}(e^{\frac{i\theta}{2}}\lambda,\dots),
\end{equation}
but a more practical way of evaluating the sum is given by the following integral:
\begin{equation}
    \sum_{h \in \mathbb{Z}}\mathcal{A}_{3,h}^{(L)}(\lambda,\dots)\mathcal{A}_{3,-h}^{(R)}(\lambda,\dots)=\frac{1}{2\pi}\int_{0}^{2\pi}{\rm d}\theta \,\mathcal{A}_{3}^{(L)}(e^{\frac{i\theta}{2}}\lambda,\dots)\mathcal{A}_{3}^{(R)}(e^{\frac{i\theta}{2}}\lambda,\dots),
\end{equation}
which can be interpreted as the statement that the trace over the Hilbert space of the intermediate state can be also evaluated in the $\theta$-basis, namely the basis of $ISO(2)$ translation eigenstates.

By splitting the three-point functions using the factorization~\eqref{eq:amplitude_CSP_helicities}, the integral over $\theta$ can be performed so that the residue of the four-point function is expressed only in terms of the infinite sum over helicity amplitudes that make the exchanged continuous-spin particle. 

Taking the external particles as massive for simplicity, the trace over the exchanged particle's Hilbert space in the $\theta$ basis can be written in terms of the factorization in eq.~\eqref{eq:amplitude_CSP_helicities} as
\begin{equation}
\begin{split}
\underset{s \to 0}{\mathrm{Res}}\,\mathcal{A}_4=\int_{0}^{2\pi}& \frac{{\rm d}\theta}{2\pi}\,  e^{-\tfrac{i\mu}{2}\left((r^{(L)}+r^{(R)})e^{i \theta}+(\tilde{r}^{(L)}+\tilde{r}^{(R)})e^{-i \theta}\right)} \\ &\times\sum_{h_1, h_2 \in \mathbb{Z}}e^{-i(h_1+h_2)\theta}\mathcal{A}_{J_1J_2h_1}^{(L)}(1^{J_1},2^{J_2},X^{h_1})\mathcal{A}_{h_2J_3J_4}^{(R)}(X^{h_2},3^{J_3},4^{J_4}),
\end{split}
\end{equation}
where $r^{(L/R)}$ correspond to the ratios associated to the left/right three-point amplitudes, and we used the homogeneity in $\lambda$ of the definite helicity components as well as that of the ratios.

Using the generating function for the Bessel function shown in eq.~\eqref{eq:Bessel} and applying the same logic as in that case, the integral can be expressed as
\begin{eqnarray}
\underset{s \to 0}{\mathrm{Res}}\,\mathcal{A}_4 &=&\int_{0}^{2\pi}\frac{{\rm d}\theta}{2\pi}  \sum_{n \in \mathbb{Z}} J_{n}(\tfrac{\mu}{2}R) \left(\frac{r^{(L)}+r^{(R)}}{i} \right)^n e^{-i n \theta} \\ && \times \sum_{h_1, h_2 \in \mathbb{Z}}e^{-i(h_1+h_2)\theta}\mathcal{A}_{J_1J_2h_1}^{(L)}(1^{J_1},2^{J_2},X^{h_1})\mathcal{A}_{h_2J_3J_4}^{(R)}(X^{h_2},3^{J_3},4^{J_4}),\quad {\rm with}\, R:=\tfrac{r^{(L)}+r^{(R)}}{\tilde{r}^{(L)}+\tilde{r}^{(R)}},\nonumber
\end{eqnarray}
which can be integrated to give
\begin{align}
    \underset{s \to 0}{\mathrm{Res}}\,\mathcal{A}_4 & = \sum_{h_1, h_2 \in \mathbb{Z}}J_{-h_1-h_2}\left(\tfrac{\mu}{2}R\right) \left(\frac{i}{r^{(L)}+r^{(R)}} \right)^{h_1+h_2} \\ \nonumber 
    & \hspace{120pt} \times \mathcal{A}_{J_1J_2h_1}^{(L)}(1^{J_1},2^{J_2},X^{h_1})\mathcal{A}_{h_2J_3J_4}^{(R)}(X^{h_2},3^{J_3},4^{J_4})\,.
\end{align}

A relevant remark is that while the contribution to a four point amplitude arising from this gluing procedure will in general {\it not} factorize into an exponential containing the continuous-spin scale $\mu$ of the exchanged particle times a function of momentum spinor brackets only, there is no tension with the general solution~\eqref{eq:amplitude_CSP}: This factorized form is what we require from amplitudes involving continuous-spin particles as external legs. If the CSP enters the amplitude as an intermediate state, there is a priori no restriction on the dependence on the amplitude on the continuous-spin scale associated to the exchanged CSP.

%*********************************************************%
\subsubsection{Some distributional higher-point amplitudes}
%*********************************************************%

There is at least one lesson from the three-point amplitudes classified above. To start, the problem is that we have to cancel the additional piece of the Lorentz generators
\begin{align}\label{additionalP}
     \sum_{k}
    \frac{\mu_k }{\langle\lambda^{(k)}\xi\rangle^2}\,
    \xi_\alpha\xi_\beta\,,
\end{align}
which every continuous-spin particles is accompanied by. There are two ways to achieve this. The first option is that the particles participating in the scattering can provide two linearly independent spinors. For example, the canonical Lorentz generator $L_{\alpha\beta}$, when acting on $\xi^\beta \psi_\beta$, yields $\sim \xi_{(\alpha} \psi_{\beta)}$, if $\psi$ transforms canonically, i.e. $\psi \neq \xi$ and is some combination of other momenta (for massive particles) and momenta spinors (for massless ones). If we have at least two such $\psi$s, the canonical piece of the Lorentz generators will produce
\begin{align}
    \sum_i \xi_{(\alpha} \psi^i_{\beta)} \frac{\pl}{\pl \langle \xi \psi^i \rangle }
\end{align}
With at least two $\phi^i$ it is should be possible to cancel the additional piece \eqref{additionalP}. We say ``should'' as there might still be further integrability conditions, see e.g. the case of one CSP particle and two massive particles of equal mass, where the Lorentz invariance in both the (anti)-holomorphic sectors leads to an additional kinematical restriction.  

The second option is realized for the case of massless fields only and for the collinear configuration therein. For (half)-collinear momenta, all $\langle ij\rangle=0$ and we do not have two linearly independent spinors to fight against the $\xi_\alpha$ contribution \eqref{additionalP}. The only way for the additional piece to cancel is to set it to zero by hand, which constrains the kinematics further. Even though the collinear configurations are not the only possibility for massless particles, generic momenta configurations may be forbidden by dynamics or other arguments. Indeed, one can think of a CSP particle as of a deformation of an infinite multiplet of massless helicity fields, for which collinear configurations seem the only possibility. It may well be that continuous-spin particles inherit this feature from the higher-spin multiplets.

With these arguments in mind, let us build some $n$-point examples. Let us assume that the additional piece \eqref{additionalP} should vanish independently. If the configuration is not collinear, we can choose $|\xi\rangle=(1,z)$, $|i\rangle=(1,z_i)$ to find that the vanishing of \eqref{additionalP} implies
\begin{align}
    &\sum_i \frac{\mu_i}{(z-z_i)^2}=0 && \Longleftrightarrow&& 
    \pl_z F(z)=0\,, &\text{with}&\quad F(z):=\sum_i \frac{\mu_i}{z-z_i}\,.
\end{align}
If we consider this as an equation with respect to $\xi$, then it corresponds to the critical points of $F(z)$. The condition is somewhat similar to the scattering equation
\begin{align}
    \sum_{i<j} \frac{k_i\cdot k_j}{z_i-z_j}&=0\,.
\end{align}
It is not difficult to find a worldsheet realization. Indeed, let us take $J(z)=\pl\phi(z)$, where $\phi(z)$ is a free boson, and consider the standard vertex operators $V_\mu = :e^{\mu \phi}:\,$. Since the operator product expansion gives
\begin{align}
    J(z)V(w) \sim \frac{\mu}{z-w}V(w)\,,
\end{align}
the function $F(z)$ is a normalized correlator $\big\langle J(z) V_{\mu_1}(z_1) \dots V_{\mu_n}(z_n)\big\rangle $. This basically means that there is a source $\sum_i \mu_i\phi(z_i)$ in the path integral. For collinear configurations, we can choose $|\xi\rangle=(1,0)$ and $|i\rangle=t_i(w,1)$ to find 
\begin{align}\label{collconstr}
    \sum_i \frac{\mu_i}{t_i^2}=0\,.
\end{align}
Note that the collinear fractions $t_i$ are Lorentz-invariant. Therefore, the following $n$-point amplitude is at least Lorentz invariant (we pay attention to the holomorphic part, while the anti-holomorphic can either be adjusted to match the holomorphic or be different if the actions on left and right are different, i.e. if the solution is chiral):
\begin{align}
    \prod_{i=1}^{n-1} \delta\big(\langle in\rangle\big)\, \lrangle{rn}^2\, \delta^{(2)}\left(\sum_j \lrangle{rj}\,[\bar{j}|_{\dot\alpha}\right) \mathcal{A}\left(\tfrac{\lrangle{ri}}{\lrangle{rn}}, [\bar{i}\,\bar{j}]\right) \delta\left(\sum_i \frac{\mu_i}{t_i^2}\right)\,.
\end{align}
Here, $r$ is an arbitrary reference spinor. The first factor imposes (half)-collinearity. The subamplitude $\mathcal{A}$ depends on Lorentz-invariant quantities. The last factor makes the additional piece \eqref{additionalP} disappear. This factor can be replaced by a more conventional
\begin{align}
    \delta\left(\sum_i \mu_i\frac{ \lrangle{\xi n}^2}{\lrangle{\xi i}^2}\right)\,,
\end{align}
where the numerator $\xi_\alpha\xi_\beta$ of \eqref{additionalP} was contracted with $n^\alpha n^\beta$, the momentum spinors of the $n$-th particle.
Another interesting option suggested by the higher-spin story is to impose ``super-collinearity''
\begin{align}
    \prod_{i=1}^{n-1} \delta^{(2)}\big(|i\rangle_\alpha-|n\rangle_\alpha\big)\, \delta^{(2)}\Big(\sum_j[\bar{j}\rvert_{\dot\alpha}\Big) \mathcal{A}\big( [\bar{i}\bar{j}]\big) \delta_{0,\sum_i \mu_i}\,.
\end{align}
Since all holomorphic spinors are equal, the constraint \eqref{collconstr} on the collinear fractions implies that the sum of continuous-spin scales vanishes. The latter, from the point of view of the two-dimensional conformal field theory, can be understood as the usual neutrality condition (which states that a correlation function of a product of vertex operators of the form $\langle\, :e^{\mu_1 \phi}: \dots :e^{\mu_n \phi}:\,\rangle$ vanishes, unless $\sum_i\mu_i=0$).

%%%%%%%%%%%%%%%%%%%%%%%%%%%%%%%%%%%%
\section{Conclusions and Discussion}\label{concl}
%%%%%%%%%%%%%%%%%%%%%%%%%%%%%%%%%%%%
We presented an extension of the spinor-helicity formalism for continuous-spin particles which,  basically, consists in relaxing the homogeneity constraint of usual massless particles when one considers continuous-spin ones. This spinor-helicity formalism allows to easily write down Lorentz-invariant on-shell amplitudes of generic massless particles, whether helicity or continuous-spin types. 

Thanks to the simplicity of the Lorentz generators for continuous-spin particles, the classification of three-point amplitudes can be approached systematically to make sure all solutions are found. Our list is in agreement with the results in the light-cone gauge obtained by Metsaev \cite{Metsaev:2017cuz,Metsaev:2018moa,Metsaev:2025wcv} and, hence, contains slightly more solutions than the list of \cite{Bellazzini:2024dco}. There are two types of solutions: (F) functional solutions, i.e. the ones that do not constrain the kinematics beyond momentum conservation and (D) distributional solutions, i.e. those that impose additional restrictions on momenta. At the end of the day, whether or not a particular three-point amplitude belongs to (F) or (D) classes is determined by the number of linearly independent spinors: if there are two, one can expand the reference spinor $\xi_\alpha$ to cancel the additional term in the Lorentz generators; if there is just one spinor in the problem (three massless particles), the additional term has to vanish by itself, which is a constraint on the kinematics of the process under consideration. 

One important example of a distributional solution (D) is the gravitational coupling, i.e. a vertex of type CSP-CSP-graviton. This coupling is not unique; more precisely, it is parameterized by a function $\mathscr{A}\left(t, [ij]\right)$ of the collinear fraction $t$, say $t_{1}$ or $t_2$ (just one of them, given the additional kinematical constraint) and by the antiholomorphic spinors, where the third leg, which belongs to the graviton, must have the right little group scaling.

We have also shown that, at least for vertices that exist for generic momenta configurations, CSP external legs of amplitudes can be obtained as the simultaneous infinite-spin and zero-mass limit from families of amplitudes where each CSP leg is replaced by a massive one. The limit explains the appearance of the exponential prefactors. It also explains the origin of the coefficients which are present in the exponential, via \eqref{eq:amp_ratios}, as fractions with which other particles' momenta are contracted with the massive spinor $\lambda^I$ that belongs to the leg on which we take the limit. It would be interesting to see if the special, distributional, solutions can also be obtained as a double-scaling limit of zero mass and infinite spin.

Perhaps, one general lesson is that continuous-spin particles favor additional constraints on the kinematics, which is very similar to the behavior of infinite multiplets of massless higher-spin fields in chiral higher-spin gravity (and its truncations, such as the higher-spin extensions of self-dual Yang-Mills and self-dual gravity). In the higher-spin case, this is clearly due to the fact that the theories have extended symmetries and are integrable \cite{Ponomarev:2017nrr}. It is not clear at the moment what kind of extended symmetries can be behind hypothetical theories with continuous-spin particles. For example, since the $\mu\to0$ limit of a CSP gives the standard higher-spin multiplet (i.e. an infinite tower of usual massless particles with all integer or half-integer helicities), it would be interesting to see if there exists a deformation of any of the known higher-spin gravity theories that makes them into theories of interacting continuous-spin particles. An encouraging feature is the existence of some non-trivial $n$-point amplitudes for continuous-spin particles. Another question in the same spirit is whether there exists a CSP analogue of the Flato--Fronsdal theorem \cite{Flato:1978qz}, i.e. whether there exists a representation $V$ (possibly non-unitary) of the Poincar\'e group (or a deformation thereof) such that the tensor product $V\otimes V$ gives a single CSP particle. For the standard massless higher-spin multiplet, such a representation $V$ of a certain chiral deformation
of the Poincaré algebra does exist \cite{Ponomarev:2022qkx}.

%%%%%%%%%%%%%%%%%%%%%%%%%
\section*{Acknowledgment}
%%%%%%%%%%%%%%%%%%%%%%%%%
We are grateful to Yannick Herfray and Laura Donnay for many useful discussions and to Ruslan Metsaev for a correspondence regarding his results in the light-cone gauge approach. F.F. is also grateful to Brando Bellazzini, Stefano De Angelis and Marcello Romano for useful comments. E.S. is a Research Associate of the Fund for Scientific Research -- FNRS, Belgium. The work of T.B., F.F. and E. S.  was partially supported by the European Research Council (ERC) under the European Union’s Horizon 2020 research and innovation programme (grant agreement No 101002551).

\appendix
%%%%%%%%%%%%%%%%%%%%%%%
\section{Miscellaneous}
\label{app}
%%%%%%%%%%%%%%%%%%%%%%%
%=========================================%
\paragraph{Comparison with the literature.}
%=========================================%
The realization of continuous-spin particles we are using here
is directly related to the one introduced in \cite{Bellazzini:2024dco}.
Indeed, we started with a pair of spinors $(\lambda,\rho)$
used to parameterise an matrix of $SL(2,\mathbb{C})$,
which can be identified  with the spinors $(\lambda,\rho)$
in the aforementioned article, up to normalization (as we require
$\langle\lambda\rho\rangle=1$), and a rescaling by a factor of $i$.
This is confirmed by the transformation law under $ISO(2)$
given in \cite[Eq. (3.7)]{Bellazzini:2024dco}, which are the same
as our right action of $ISO(2)$  on the matrix \eqref{matrixformed} formed by $\lambda$ and $\rho$,
upon taking into account the imaginary unit factor that is a matter
of convention.

Consequently, our amplitudes should be related to those
of \cite{Bellazzini:2024dco} by applying the isomorphism 
\eqref{eq:isomorphism} on each leg with a continuous-spin particle,
i.e. we should compare
\begin{equation}
    \prod_{k \in I_{\scriptscriptstyle\sf CSP}}
    \exp\bigg(-\tfrac{i\mu_k}2\,\Big[\tfrac{\langle\rho^{(k)}\xi\rangle}{\langle\lambda^{(k)}\xi\rangle}e^{-i\theta_k}
    +\tfrac{[\tilde\rho^{(k)}\tilde\xi]}{[\tilde\lambda^{(k)}\tilde\xi]}e^{i\theta_k}\Big]\bigg)\,
    \mathcal{A}_\xi\big(e^{i\theta_i/2}\lambda^{(i)},\dots\big)
\end{equation}
where the dots denote momentum spinors other
than the CSP ones,
with the results of \cite{Bellazzini:2024dco}.
First, note that since the ratio $r_k$ is homogeneous
of degree $-2$ in $\lambda^{(k)}$, it verifies
\begin{equation}
    r_k\big(e^{i\theta_k/2}\lambda^{(k)},\psi^{(m)}\big)
    = e^{-i\theta_k}\,r_k\big(\lambda^{(k)},\psi^{(m)}\big)\,,
\end{equation}
and the same is true of $\tilde r_k$
with respect to $\tilde\lambda^{(k)}$. On top of that,
\begin{align}
    \tfrac{\langle\rho^{(k)}\xi\rangle}{\langle\lambda^{(k)}\xi\rangle} + r_k
    = \tfrac1{\langle\lambda^{(k)}\xi\rangle}
    \sum_i a_{k,i} \Big[\langle\rho^{(k)}\xi\rangle
    +\tfrac{\langle\xi\psi^{(i)}\rangle}{\langle\lambda^{(k)}\psi^{(i)}\rangle}\Big]
    = \sum_i a_{k,i}\frac{\langle\rho^{(k)}\psi^{(i)}\rangle}
    {\langle\lambda^{(k)}\psi^{(i)}\rangle}\,,
\end{align}
upon using the fact that the $a_{k,i}$ sum to $1$
for any fixed value of $k$, and the Schouten identity once more,
so that the transformed amplitude reads
\begin{equation}\label{CSPampli}
    \prod_{k \in I_{\scriptscriptstyle\sf CSP}}
    \exp\bigg(-\tfrac{i\mu_k}2\,
    \Big[e^{-i\theta_k}\sum_i a_{k,i}\tfrac{\langle\rho^{(k)}\psi^{(i)}\rangle}
    {\langle\lambda^{(k)}\psi^{(i)}\rangle}
    + e^{i\theta_k}\sum_j \tilde a_{k,j}\tfrac{[\tilde\rho^{(k)}\tilde\psi^{(i)}]}
    {[\tilde\lambda^{(k)}\tilde\psi^{(i)}]}\Big]\bigg)\,
    \mathscr{A}\big(e^{i\theta_i/2}\lambda^{(i)},\dots\big)\,,
\end{equation}
which agrees with the general form of $n$-pt amplitudes
derived in \cite{Bellazzini:2024dco} (again, up to a factor 
of the imaginary unit $i$ that is due to a convention difference
in our definition of the spinor $\rho$).

%==========================%
\paragraph{Patching things.}
%==========================%
Given $\xi\in\mathbb{C}^2\backslash\{0\}$, we have seen
that an $ISO(2)$-equivariant function on $SL(2,\mathbb{C})$-valued in functions on the circle, in the patch
\begin{equation}
    \mathbb{C}^2\backslash\{0\} \supset U_\xi
    := \big\{\lambda\in\mathbb{C}^2\backslash\{0\}
    \mid \langle\lambda\xi\rangle\neq0\big\}\,,
\end{equation}
can be written as
\begin{equation}
    F(\lambda,\rho,\theta)\rvert_{U_\xi}
    = e^{-i\mu\,\Re\Big[\frac{\langle\rho\xi\rangle}{\langle\lambda\xi\rangle}e^{-i\theta}\Big]}
    \Phi_\xi(e^{i\theta/2}\lambda)\,.
\end{equation}
In order to cover all of the $\mathbb{C}^2\backslash\{0\}$ space
allowed for $\lambda$, we only need one other patch, say $U_\eta$
defined by $\eta\in\mathbb{C}^2\backslash\{0\}$
linearly independent from $\xi$, i.e. $\langle\xi\eta\rangle\neq0$.
On the overlap of these two patches, we have
\begin{equation}
    F(\lambda,\rho,\theta)\rvert_{U_\xi \cap U_\eta}
    = e^{-i\mu\,\Re\Big[\frac{\langle\rho\xi\rangle}{\langle\lambda\xi\rangle}e^{-i\theta}\Big]}
    \Phi_\xi(e^{i\theta/2}\lambda)
    = e^{-i\mu\,\Re\Big[\frac{\langle\rho\eta\rangle}{\langle\lambda\eta\rangle}e^{-i\theta}\Big]}
    \Phi_\eta(e^{i\theta/2}\lambda)\,.
\end{equation}
which imposes that
\begin{align}\label{phasepatch}
    \Phi_\eta(\lambda) 
    = g_{\xi\eta}(\lambda)\Phi_\xi(\lambda)\,,
    \qquad 
    g_{\xi\eta}(\lambda) := e^{i\mu\,\Re\Big[\frac{\langle\xi\eta\rangle}{\langle\lambda\xi\rangle\langle\lambda\eta\rangle}\Big]}\,,
\end{align}
where we used $\langle\lambda\rho\rangle=1$ and took advantage of the fact that the dependency 
on $\theta$ can be absorbed completely in a phase of $\lambda$.
The phase in \eqref{phasepatch} can be interpreted as a $U(1)$-valued function
defined on the overlap of two patches 
\begin{equation}
    g_{\xi\eta}: U_\xi \cap U_\eta \longrightarrow U(1)\,,
\end{equation}
which verifies\footnote{The second identity again follows
from the Schouten identity.}
\begin{equation}
    g_{\xi\xi}(\lambda) = 1\,,
    \qquad 
    g_{\xi\eta}(\lambda)g_{\eta\zeta}(\lambda)g_{\zeta\xi}(\lambda) = 1\,,
\end{equation}
i.e. they define \emph{transition functions}
for a \emph{complex line bundle} with $U(1)$ structure group,%
\footnote{Let us remark that these transition functions
\emph{factorise}, in the sense that they that the form
$g_{\xi\eta}=t_\xi t_\eta^\inv$,
with $t_\xi = \exp\big[-i\mu\,\Re\big(\frac{\langle\rho\xi\rangle}{\langle\lambda\xi\rangle}\big)\,\big]$. Transitions functions of
a vector bundle are \v{C}ech cocycles in general,
with values in the Abelian circle group $U(1)$ in our case,
and the fact that they admit the aforementioned factorised form
amounts to saying that they are \emph{exact}. So this complex line bundle is \emph{trivial}. In fact, the equation \eqref{eq:u} provides an explicit trivialisation.}
whose local sections are our continuous-spin fields.
Note that for $\eta$ and $\xi$ infinitesimally close, in the sense
\begin{equation}
    \eta = \xi + \varepsilon\,\delta\xi\,,
\end{equation}
with $\delta\xi$ understood as a deviation from $\xi$
(in particular it is not collinear to it) and $\varepsilon$
an infinitesimal parameter, then one finds
\begin{equation}
    \Phi_{\eta} = \Phi_\xi + \varepsilon\,\tfrac{i\mu}{2}\Big(\tfrac{\langle\xi\delta\xi\rangle}{\langle\lambda\xi\rangle^2} + \tfrac{[\bar\xi\delta\bar\xi]}{[\bar\lambda\bar\xi]^2}\Big)\,\Phi_\xi + \mathcal{O}(\varepsilon^2)\,.
\end{equation}

\paragraph{Massless limit.}
%=========================%
Let us recall that, in the standard spinor-helicity
representations for massive fields,
one uses a pair of two-component spinors to parameterise
massive momenta, and consider this pair as an $SU(2)$-doublet. This pair of spinors can be thought of,
again, as parameterising an $SL(2,\mathbb{C})$ matrix,
and the wavefunction for massive fields is obtained
as a function of this pair of spinors, 
taking values in the spin-$J$ irreducible representation of $SU(2)$,
with some equivariance property. More precisely,
consider $F(\lambda_\alpha^I,\tilde\lambda_{K\,\dot\alpha},u_L)$
with $u_I$ an $SU(2)$-doublet (with $I,K,L=1,2$) used to realise 
the $(2J+1)$-dimensional representation as symmetric
rank-$2J$ tensors, in the sense
\begin{equation}\label{homogeneitymassive}
    F(\lambda,\tilde\lambda,tu) 
    = t^{2J}\,F(\lambda,\tilde\lambda,u)\,.
\end{equation}
The condition that $F$ be equivariant under the action
of $SU(2)$ can be concisely written as
\begin{equation}\label{massivequiv}
    \Big(\lambda^I_\alpha\,\frac{\partial}{\partial\lambda_\alpha^J}
    - \tilde\lambda_{J\,\dot\alpha}
    \frac{\partial}{\partial\tilde\lambda_{I\,\dot\alpha}}
    - u_J\,\frac{\partial}{\partial u_I}
    -\frac12\,\delta^I_J\,\mathscr{E}\Big)
    F(\lambda,\tilde\lambda,u) = 0\,,
\end{equation}
where the Euler vector field $\mathscr{E}$ is defined as
\begin{equation}
    \mathscr{E} := \lambda^I_\alpha\,\frac{\partial}{\partial\lambda_\alpha^I}
    - \tilde\lambda_{I\,\dot\alpha}
    \frac{\partial}{\partial\tilde\lambda_{I\,\dot\alpha}}
    - u_I\,\frac{\partial}{\partial u_I}\,,
\end{equation}
and the above condition \eqref{massivequiv} is simply solved by any function of the $SU(2)$-invariants
\begin{equation}\label{eq:su2_inv}
    p_{\alpha\dot\alpha} = \lambda_\alpha^I\tilde\lambda_{I\,\dot\alpha}\,,
    \qquad 
    \zeta_\alpha := u_I\lambda^I_\alpha\,,
    \qquad 
    \tilde\zeta_{\dot\alpha} = u^I\,\tilde\lambda_{I\,\dot\alpha}\,.
\end{equation}
Noticing that, upon fixing the angle brackets 
of the momentum spinors as
\begin{equation}
    \langle\lambda^I\lambda^J\rangle = m\,\epsilon^{IJ}\,,
\end{equation}
these three types of invariants are related via
\begin{equation}\label{eq:rel_psi}
    m\,\tilde\zeta_{\dot\alpha} = p_{\alpha\dot\alpha}\,\zeta^\alpha\,.
\end{equation}
Therefore, one can write the solution of \eqref{homogeneitymassive}-\eqref{massivequiv} as
\begin{equation}\label{eq:massive}
    F(\lambda,\tilde\lambda,u) = \zeta_{\alpha_1} \dots \zeta_{\alpha_{2J}}\,
    F^{\alpha_1 \dots \alpha_{2J}}(p_{\beta\dot\beta})\,,
\end{equation}
without loss of the generality.
Next, we want to discuss how to take the massless limit in order
to recover massless particles of helicity type,
as well as of continuous-spin type.

Let us first discuss the contraction of finite-dimensional
$SU(2)$-representations to the finite-dimensional but unfaithful 
irreducible representations of $ISO(2)$, as well as the infinite-dimensional and faithful ones.
The fundamental representation of $SU(2)$ is $\mathbb{C}^2$,
with the action simply given by 
\begin{equation}
    \begin{pmatrix}
        a & b \\
        -m^2\bar b & \bar a
    \end{pmatrix}
    \begin{pmatrix}
        u_1 \\ u_2
    \end{pmatrix}
    = \begin{pmatrix}
        a\,u_1+b\,u_2 \\
        -m^2\bar b\,u_1 + \bar a\,u_2
    \end{pmatrix}\,,
    \qquad\qquad |a|^2 + m^2|b|=1\,,
\end{equation}
with $m\to0$ implementing the contraction $SU(2) \to ISO(2)$.
The $(2J+1)$-dimensional irreducible representations, with $J \in \tfrac12\mathbb{N}$,
which are obtained as the symmetric tensor product of the fundamental
representation $2J$ times, can be realised as polynomials 
in $u_{I}$ which are homogeneous of degree $2J$.
% \begin{equation}
%     a = 1 + \varepsilon\,i\alpha + \mathcal{O}(\varepsilon^2)\,,
%     \qquad
%     b = \varepsilon\,\beta + \mathcal{O}(\varepsilon^2)\,,
%     \qquad 
%     \alpha \in \mathbb{R}\,,
%     \quad \beta\in\mathbb{C}\,
% \end{equation}
Redefining $u_2 \to m\,u_2$, the Lie algebra action is given 
on this representation by
\begin{equation}
    L := \tfrac12\big(u_1\partial_{u_1} - u_2\partial_{u_2}\big)\,,
    \qquad 
    P_- = m\,u_2\partial_{u_1}\,,
    \qquad 
    P_+ = m\,u_1\partial_{u_2}\,,
\end{equation}
which makes it clear that, in the contraction limit $m\to0$,
the action of the translation generators of $ISO(2)$, i.e. $P_\pm$, becomes trivial.
One is therefore left with the same vector space, whose basis
can be chosen as monomials of the form $u_1^{J+k}u_2^{J-k}$,
which are eigenvector of $L$ with eigenvalue $k \in \{-J,\dots,J\}$,
and therefore define one-dimensional irreducible representations of
$U(1) \cong SO(2) \subset ISO(2)$.

In order to work out the contraction to a faithful representation
of $ISO(2)$, let us re-write homogeneous polynomials
on $\mathbb{C}^2$ as
\begin{equation}
    f(u_I) = (u_1u_2)^J\,p_J(\tfrac{u_1}{u_2})\,,
\end{equation}
where $p_J$ is a Fourier series (or, rather, sum) with non-zero modes 
$-J \leqslant n \leqslant J$. One can therefore transfer 
the action of $\mathfrak{su}(2)$ on polynomials
of one complex variable $z=\tfrac{u_1}{u_2}$ via
\begin{equation}
    (u_1u_2)^{-J} L\, (u_1u_2)^J = z\,\partial_z\,,
    \qquad 
    (u_1u_2)^{-J} P_\pm \,(u_1u_2)^J = m\,z^{\pm1}\,(J \mp z\partial_z)\,
    \underset{\substack{m\to0\\ J\to\infty}}{\longrightarrow}\,
    \mu\,z^{\pm1}\,,
\end{equation}
which yields the faithful irreducible representation of $ISO(2)$ in the double-scaling limit
$m \to 0$ and $J \to \infty$ with parameter $\mu=mJ>0$ kept fixed,
upon evaluating $z$ on the circle.

Starting from the form \eqref{eq:massive} of the massive
spin-$J$ wavefunctions,
\begin{equation}
    F^J(\lambda,\tilde\lambda,u) = \zeta_{\alpha_1} \dots \zeta_{\alpha_{2J}}\,
    F^{\alpha_1 \dots \alpha_{2J}}(p)\,,
    \qquad 
    \zeta_\alpha = u_I\lambda^I_\alpha\,,
    \qquad 
    p_{\alpha\dot\alpha} = \lambda^I_\alpha\tilde\lambda_{I\,\dot\alpha}\,,
\end{equation}
let us re-write it as
\begin{equation}\label{eq:prep}
    F^{\alpha_1 \dots \alpha_{2J}}(p) \zeta_{\alpha_1} \dots \zeta_{\alpha_{2J}}
    = (-\tfrac{1}{m})^J F^{\alpha_1 \dots \alpha_{2J}}(p) 
    \zeta_{\alpha_1} \dots \zeta_{\alpha_J} p_{\alpha_{J+1}}{}^{\dot\alpha_1}
    \dots p_{\alpha_{2J}}{}^{\dot\alpha_J}
    \tilde\zeta_{\dot\alpha_1} \dots \tilde\zeta_{\dot\alpha_J}\,,
\end{equation}
upon using the relation \eqref{eq:rel_psi} between
the $SU(2)$ invariants $\zeta_\alpha$ and $\tilde\zeta_{\dot\alpha}$.
Next, introduce $\xi \in \mathbb{C}^2\backslash\{0\}$
with $\langle\lambda\xi\rangle\neq0$ so that one can decompose
$\lambda^{I=2} \equiv m\rho$ in terms of $\lambda^{I=1}\equiv\lambda$
and $\xi$ as
\begin{equation}
    \rho_\alpha = \frac{\langle\rho\xi\rangle}{\langle\lambda\xi\rangle}\,\lambda_\alpha 
    + \frac{1}{\langle\lambda\xi\rangle}\,\xi_\alpha\,.
\end{equation}
\begin{subequations}
Plugging this decomposition in the $SU(2)$ invariants,
and setting $u_1=e^{i\theta/2}$ and $u_2=ie^{-i\theta/2}$, yields
\begin{align}
    \zeta_\alpha & = \Big(1+im\,e^{-i\theta}
    \tfrac{\langle\rho\xi\rangle}{\langle\lambda\xi\rangle}\Big)
    e^{i\theta/2}\lambda_\alpha 
    + im\,e^{-i\theta/2}\,\frac{1}{\langle\lambda\xi\rangle}\,\xi_\alpha\,, \\
    -i\tilde\zeta_{\dot\alpha} & = \Big(1+im\,e^{i\theta}
    \tfrac{[\tilde\rho\tilde\xi]}{[\tilde\lambda\tilde\xi]}\Big)
    e^{-i\theta/2}\,\tilde\lambda_{\dot\alpha} 
    + im\,e^{i\theta/2}\,\frac{1}{[\tilde\lambda\tilde\xi]}\,\tilde\xi_{\dot\alpha}\,,
\end{align}
\end{subequations}
and one can therefore re-express 
the massive spin-$J$ wavefunctions as
\begin{equation}
    F^J_\xi(\lambda,z) = \Big(1+im\,e^{-i\theta}
    \tfrac{\langle\rho\xi\rangle}{\langle\lambda\xi\rangle}\Big)^J
    \Big(1+im\,e^{i\theta}\tfrac{[\tilde\rho\tilde\xi]}{[\tilde\lambda\tilde\xi]}\Big)^J\,
    \times \sum_{h=-J}^{J} C_\xi^h(e^{i\theta/2}\,\lambda)\,,
\end{equation}
where we have defined $C_\xi^h$ to be the component of degree $2h$
in $\lambda$, whose components are obtained from those 
of $F^{\alpha_1 \dots \alpha_{2J}}$ contracting them with various powers
of $\xi_\alpha$ and $\tilde\xi_{\dot\alpha}$ (rescaled by a power 
of the mass), respectively corresponding to the helicity $h$, $0$ and $-h$,
for $1 \leqslant h \leqslant J$, of a massive field. Taking the massless limit
can be done on wavefunctions  by sending $m\to0$
so that the massive momentum parameterised by the spinors $\lambda$ and $\rho$
becomes a massless momentum determined by $\lambda$ only,
and by considering the above spinor-tensors as independent functions
of the momentum (i.e. forgetting about the mass-dependent 
redefinition made in the three equations above). Each of these 
helicity components becomes independent and thus we recover
the expected massless limit of a massive field of spin-$J$
to massless fields of helicity $-J \leqslant h \leqslant J$.

In doing so, the prefactor also disappears. However, if we take
a slightly more refined limit, consisting in send $m\to0$
while simultaneously sending $J\to\infty$, so that 
upon changing $m \leadsto \tfrac{\mu}{2J}$, one finds
\begin{equation}
    \Big(1+im\,e^{-i\theta}
    \tfrac{\langle\rho\xi\rangle}{\langle\lambda\xi\rangle}\Big)^J
    \Big(1+im\,e^{-i\theta}\tfrac{[\tilde\rho\tilde\xi]}{[\tilde\lambda\tilde\xi]}\Big)^J 
    \underset{\substack{m\to0\\ J\to\infty}}{\longrightarrow}\
    \exp\Big(i\mu\Re\Big[e^{-i\theta}
    \tfrac{\langle\rho\xi\rangle}{\langle\lambda\xi\rangle}\Big]\Big)\,,
\end{equation}
and therefore the full massive spin-$J$ wavefunction
becomes, in this limit,
\begin{equation}
    F^J(\lambda,\tilde\lambda,u) 
    \underset{\substack{m\to0\\ J\to\infty}}{\longrightarrow}\
    \exp\Big(i\mu\Re\Big[e^{-i\theta}
    \tfrac{\langle\rho\xi\rangle}{\langle\lambda\xi\rangle}\Big]\Big)
    \Phi_\xi(e^{i\theta/2}\lambda)\,,
\end{equation}
with $\Phi_\xi$ being a function with no fixed homogeneity 
in $\lambda$, or in other words, containing all helicities,
as expected of a continuous-spin wavefunction.

Now let us try and apply this to amplitudes involving massive
and massless particles of helicity-type. Lorentz invariance
implies that, up to a momentum-conserving Dirac delta distribution,
this is a function of the angle- and square- brackets 
of the various momentum spinors. For massless legs,
helicity $h$ is fixed by requiring the amplitude to be homogeneous 
of degree $-2h$, and similarly, for massive legs, spin is fixed
by requiring the amplitude to be homogeneous of degree $2J$
in the combinations $\zeta$ and $\tilde\zeta$
defined in \eqref{eq:su2_inv}. So schematically, 
\begin{equation}
    \mathcal{A}_{\{J_i\}, \{h_j\}} = \delta^{(4)}\Big(\sum_k p_k\Big)\,
    \times \mathscr{A}\big(\langle\zeta^{(i)}\zeta^{(j)}\rangle, 
    \langle\zeta^{(i)}\chi^{(j)}\rangle,[\tilde\zeta^{(i)}\tilde\zeta^{(j)}],
    [\tilde\zeta^{(i)}\tilde\chi^{(j)}]\dots\big)\,,
\end{equation}
where $\chi^{(i)}$ and $\tilde\chi^{(j)}$ denote the momentum spinors for the massless 
particles of helicity-type. Let us fix a massive particle
whose `continuous-spin limit' we want to take, say initially
of spin-$J$ and with $SU(2)$-invariant combination of momentum
spinors denoted by $\zeta$ as usual. The angle and square brackets
of the latter with any other spinor, generically denoted by 
the placeholder $\bullet$ below, reads
\begin{equation}
    \langle\zeta\bullet\rangle 
    = e^{i\theta/2}\langle\lambda\bullet\rangle\,
    \Big(1+ime^{-i\theta}\tfrac{\langle\rho\bullet\rangle}{\langle\lambda\bullet\rangle}\Big)\,,
    \qquad 
    -i\,[\tilde\zeta\bullet] 
    = e^{-i\theta/2}[\tilde\lambda\bullet]\,
    \Big(1+ime^{i\theta}\tfrac{[\tilde\rho\bullet]}{[\tilde\lambda\bullet]}\Big)\,,
\end{equation}
where as above we identified $u_1=e^{i\theta/2}$ 
and $u_2=ie^{-\theta/2}$, and renamed $\lambda^{I=1}=\lambda$ 
and $\lambda^{I=2}=m\rho$.
Assuming that the delta-stripped part of the amplitude
is a polynomial or rational function of the angle 
and square brackets, it can be decomposed as a sum 
\begin{equation}
    \mathscr{A} = \sum_{\sum_i w_i=J=\sum_j \tilde w_j}
    \mathscr{A}_{\{w_i,\tilde w_j\}}
\end{equation}
with
\begin{equation}
    \mathscr{A}_{\{w_i,\tilde w_j\}}\big(t_i\langle\zeta\psi^{(i)}\rangle,
    \tilde t_j[\tilde\zeta\tilde\psi^{(j)}],\dots\big)
    = \prod_i t_i^{w_i} \prod_j \tilde t_j^{\,\tilde w_i}\,\times\,
    \mathscr{A}_{\{w_i,\tilde w_j\}}\big(\langle\zeta\psi^{(i)}\rangle,
    [\tilde\zeta\tilde\psi^{(j)}],\dots\big)\,,
\end{equation}
with the set $\{w_i\}$ being integers and $\psi^{(i)}$ 
and $\tilde\psi^{(j)}$ standing for any (combinations of)
other momentum spinors than those of the massive spin-$J$ particle
we singled out. The idea is simply
that since the full $\mathscr{A}$ is homogeneous of degree $2J$
in $\zeta$, and that the latter enter only through angle 
and square brackets, the full amplitude is in general 
not homogeneous in each of these brackets, but can however
be decomposed into a sum of pieces which have such a definite
homogeneity in these brackets, whose degrees add up to $2J$.
The particular split consisting in the total homogeneity
coming from the angle brackets being the same 
as that coming from the square brackets, i.e. $J$,
is justified by the fact that the invariants $\zeta$
and $\tilde\zeta$ are not independent but related by the momentum,
and hence the massive wavefunction can be brought into
a `symmetric' form \eqref{eq:prep} as we found useful
to discuss the contraction at the one-particle level.
Now define 
\begin{equation}
    b_i := \tfrac{w_i}{J}\,,
    \quad \tilde b_i := \tfrac{\tilde w_i}{J}
    \qquad\Longrightarrow\qquad
    \sum_i b_i = 1 = \sum_j \tilde b_j\,,
\end{equation}
so that in the continuous-spin limit, 
\begin{equation}
    \mathscr{A}_{\{w_i,\tilde w_j\}}\
    \underset{\substack{m\to0\\J\to\infty}}{\longrightarrow}\
    \exp\Big(i\mu\,\sum_i b_i e^{-i\theta}\tfrac{\langle\rho\psi^{(i)}\rangle}{\langle\lambda\psi^{(i)}\rangle}
    + \tilde b_i e^{i\theta}\tfrac{[\tilde\rho\tilde\psi^{(i)}]}{[\tilde\lambda\tilde\psi^{(i)}]}\Big)\times\sum_{h\in\mathbb{Z}} \mathscr{A}_h(e^{i\theta/2}\langle\lambda\psi^{(i)}\rangle,\dots)
\end{equation}
which produces the expected exponential factor associated with
a CSP leg.

\newpage

\setstretch{1.0}
\providecommand{\href}[2]{#2}\begingroup\raggedright\endgroup

\end{document}